\documentclass[pra,aps,showkeys,twocolumn]{revtex4}
\usepackage[dvips]{graphicx}
\usepackage{subfigure} 
\begin{document}
%\draft

% define new commands
\newcommand{\wat}{H$_{2}$O\ }
\renewcommand{\subfigtopskip}{-3pt}
\renewcommand{\subfigcapskip}{-3pt}
\renewcommand{\subfigbottomskip}{-3pt}

\title{%
Temperature Dependent Conformational Transitions
and Hydrogen Bond Dynamics
of the Elastin--Like Octapeptide GVG(VPGVG):
a Molecular Dynamics Study
}

\author{%
Roger Rousseau~$^{(1),(2)}$,
Eduard Schreiner~$^{(1)}$,
Axel Kohlmeyer~$^{(1)}$,
and
Dominik Marx~$^{(1)}$
}
\affiliation{
\mbox{$^{(1)}$
Lehrstuhl f\"ur Theoretische Chemie, Ruhr--Universit\"at Bochum,
44780 Bochum, Germany}\\
\mbox{$^{(2)}$
International School for Advanced Studies, Via Beirut 4,
34014, Trieste, Italy}\\
}

\date{September 16, 2003}

\begin{abstract}
A joint experimental / theoretical investigation of the 
elastin-like octapeptide GVG(VPGVG) was carried out.
In this paper a comprehensive molecular dynamics study of the
temperature dependent folding and unfolding of the octapeptide
is presented.
The current study, as well as its experimental 
counterpart
(see previous paper)
find that this peptide undergoes an 
``inverse temperature transition'', ITT, 
leading to a folding at about 40-50$^\circ$C.
In addition, an unfolding transition is identified
at unusually high temperatures approaching the boiling point of water.
Due to the small size of the system two 
broad temperature regimes are found:
the ``ITT regime'' (at about 10-50$^\circ$C)
and the ``unfolding regime'' at about $T>60^\circ$C,
where the peptide has a maximum probability
of being folded at $T\approx 60^\circ$C.
A detailed molecular picture involving a thermodynamic
order parameter, or reaction coordinate, for this process is 
presented along with a
time-correlation function analysis of the hydrogen bond dynamics
within the peptide as well as 
between the peptide and solvating water molecules.
Correlation with experimental evidence and ramifications on 
the properties of elastin are discussed.
\end{abstract}
\maketitle

\newpage

%\twocolumn
%\narrowtext
 
%%%%%%%%%%%%%%%%%%%%%%%%%%%%%%%%%%%%%%%%%%%%%%%%%%%%%%%%%%%%%%%%%%%%%%%%%%%
%%%%%%%                       BODY OF TEXT
%%%%%%%%%%%%%%%%%%%%%%%%%%%%%%%%%%%%%%%%%%%%%%%%%%%%%%%%%%%%%%%%%%%%%%%%%%%

\section{Introduction}
\label{intro}

Vertebrate elastic fibers, as contained in vascular  
walls, skin or lungs, allow for reversible deformations upon
mechanical stress.
These fibers consist of two main types of protein 
components: a fibrous component, {\em fibrilen},
and an amorphous component, the globular {\em elastin} 
protein~\cite{ciba}.
The essential elasticity is provided by the 
latter protein, elastin,
which is known to have very unique viscoelastic 
properties in the water swollen state 
(see Refs.~\cite{urry-jpcB,tamburro99,alix99,Rees01,Tamburro02,urry02} for 
recent overviews).
The insoluble elastin 
is an extensively crosslinked polymer of 
a precursor {\em tropoelastin}.
Tropoelastin is composed of two types of domains. 
One of them is rich in lysine
and provides cross-linking of the individual 
monomers resulting in
lysinonorleucin, desmosine and isodesmosine 
links, which are
responsible for its typical yellow color.
The other is made predominantly of 
hydrophobic amino acids with the
highly repetitive pentameric repeat unit
``VPGVG'' (amino acids V: valine, P: proline, G: glycine)
which has been found to be crucial to elastin's 
functionality~\cite{urry-angew,urry-jpcB,rees98,urry02}.

An interesting peculiarity of 
both elastin and tropoelastin, is that they
undergo {\em folding} by {\em increasing} the temperature
beyond typically 25$^\circ$C.
The term ``inverse temperature transition'' (ITT) was coined for
this apparently paradoxical change from a ``disordered'' (extended)
to an ``ordered'' (folded) conformation upon heating.
A striking manifestation of this phenomena is the ability
to grow crystalline solid state
structures from a solution of cyclic elastin-like oligopeptides by 
{\em heating}~\cite{cryst1,cryst2}.
As such, tropoelastin and its synthetic analogues, 
have been a subject of intense
investigation in the field of biopolymers and protein engineering.
Elastin-like polymers have the promise of providing
materials for bio-mechanical devices~\cite{urry-angew,urry-jpcB,Chilkoti} 
as well as temperature dependent molecular switches~\cite{rees98}.
Despite the intense research efforts, the detailed
structure of tropoelastin still defeats elucidation.
Hence a molecular level picture which relates
the protein structure to the viscoelastic properties
is still a matter of debate.
What is beyond doubt, however,
is the decisive role of water as a plasticizer~--
dry elastin is {\em brittle}~\cite{Partidge62}.
Still, the potentially crucial aspect
of the protein's hydration water 
remains largely unexplored with 
the notable exception of recent MD 
simulations~\cite{daggett01}.

The origin of the viscoelastic properties
of elastin is controversially discussed,
calling concepts such as
classical rubber elasticity~\cite{flory},
various librational entropy 
mechanism~\cite{wasserman90,tamburro99,urry-angew},
and multi--phase models~\cite{alix99,daggett01}.
Much of the experimental data has been 
summarized in excellent
reviews~\cite{urry-jpcB,Tamburro02,daggett02c} 
on both elastin and elastin-like polypeptides. 
There is much data that suggest that both elastin and its synthetic mimics
display an interesting conformational dynamics in solution.
Early NMR studies suggest that elastin under physiological conditions is
composed of highly mobile chains~\cite{NMR1,NMR2},
which correlated well with the observation
from  birefringence~\cite{biref} measurements that elastin, 
in the same state, is isotropically distributed with the chains
adopting random conformations.
Studies of the thermo-mechanical properties of water swollen 
elastin and tropoelastin are consistent 
with the interpretation of
elastin as a classical rubber 
with heavy
emphasis on the role of entropic contributions 
to elasticity~\cite{flory}.
These findings would support a single phase
model where the structure of elastin is
random and undergoes large amplitude fluctuations.
Recent single-molecule AFM and spectroscopic
measurements~\cite{urry02} studies have severely challenged
this interpretation and  suggest that, above the ITT temperature, 
a structurally {\em ordered} model of
poly(VPGVG) can also account for these elastic properties
due to significant protein librational contributions to the 
overall entropy as suggested two decades ago~\cite{urry83}. 
Other spectroscopic experiments such as 
FT-IR, NMR, CD and Raman 
measurements~\cite{NMR3,urry85,rees98,joint,winter,debelle95} 
on elastin-like polypeptides
suggest the presence of $\beta$- and $\gamma$-turns
as a common structural motif 
which may be 
in dynamic equilibrium and very short and/or 
distorted antiparallel $\beta$-strands and disordered 
structures with possible dynamic 
sheet-coil-turn transitions.
Recent NMR studies also point to the high mobility of
the elastin chains 
in the water swollen elastins~\cite{perry02}
and in solution multiple temperature states for
poly(GVGVP), each with different dynamical
behavior and peptide-water interactions, 
were observed~\cite{Kurkova}.
These later data would rather support
a multiphase model of elastin~\cite{alix99}
where the fine details of the protein structure and in particular
its
interactions with water play a crucial role in understanding
its elasticity.
Overall, there is a wealth of data on the structure
and properties of these systems, and although
the models do not agree on their interpretation,
a central ingredient
is the intriguing and complex dynamical
behavior of elastin-like polymers themselves and the crucial interplay
with solvating water molecules.

Atomistic simulations on elastin or elastin-analogues are 
scarce. 
Early work (see Ref.~\cite{tamburro99} for a review),  such
as Ref.~\cite{wasserman90}, studied the elastic
properties of these proteins by performing short
MD simulations under constraints in order to mimic
an external pulling force.
Here the authors found that entropic forces from
reduced librational motion of the protein were 
indeed a main contributing factor in the elastic 
behavior.
However, the authors were understandably limited to simulation times
not exceeding one nanosecond.
It is noted that, in general, short 
simulation times generally lead to 
unreliable statistics,  due to insufficient
sampling of configuration space,
and hence may provide only {\em qualitative} insights.
A crucial factor also to be considered is the explicit
atomistic treatment 
of the protein/water interface and characterization of its role 
in determining  elastin's properties.
As a significant advance along these lines, recent 
studies on solvated 90 amino acid polypeptides of 
sequence (VPGVG)$_{18}$,
have been reported, which explicitly include 
water~\cite{daggett01}.
The importance of the protein/water interface
was worked out based on thermodynamic and
static structural -- as opposed to dynamic -- considerations.  
This work was successfully able to reproduce the ITT, 
clarify how water may contribute
to the entropic contribution to the 
elastic forces~\cite{daggett02}
as well as probe the role of elastin mimics
as potential molecular machines~\cite{daggett02b}.
Overall this work was was successfully 
able to qualitatively relate
its findings to much of the 
experimentally observed phenomena, though, some 
debate has arisen with regards to the ability of 
the proposed structures to 
account for the temperature dependence
of elastin's water content above the ITT~\cite{urry02}.
Despite the significant improvement these simulations have 
brought to our understanding 
of the atomic level details 
concerning elastins' behavior,
these quite long biopolymers were simulated 
on the order of a few nanoseconds for a given temperature.
In view of the expected long relaxation times of
proteins of that size
~\cite{urry85a} 
it would be highly desirable to have access to 
simulation times that allow for
quantitative statistical-mechanical analysis for a given 
number of amino acids.
This is especially true
for low frequency oscillations of the protein backbone 
which play a central role in many 
models of elastin's properties.
Furthermore, the {\em dynamics} and in particular
the {\em kinetics} of hydrogen-bonds remains unexplored
up to now. 
Thus, questions regarding 
the coupling of the dynamics of the protein, 
the protein/water interface
and hydrogen bonding remain open issues.

A significant experimental finding
was the recent demonstration~\cite{rees98} that
oligopeptides of the kind GVG(VPGVG)$_n$
display the ITT even in the limit 
of only one pentameric repeat unit.
Using the limiting value $n=1$ 
opens up the possibility to 
perform  MD simulations that allow
for fairly long simulation times with sufficiently many 
solvating water molecules. 
We therefore have launched a joint 
experimental~/ simulation study of
the smallest possible elastin model,
the octamer GVG(VPGVG) (having only $\approx 640$~Da), see
the preceding paper~\cite{winter}
and Ref.~\cite{joint} for a preliminary short communication.
This minimal elastin model was chosen,
at the expense of experimental
difficulties, in order to allow for extensive
molecular dynamics (MD) simulations, at 12 temperatures
between 280~K--390~K,
with a thorough conformational sampling of 32 nanoseconds each 
(after equilibration),
followed by in--depth statistical mechanical analysis 
with a focus on dynamics.
The current approach will allow us
to obtain a crucial link between the 
multitude of experimental results and 
the atomistic simulations by directly
comparing results obtained on essentially
the same system under thermodynamically 
consistent conditions.
Moreover, the detailed statistical mechanical analysis which
is possible on the current MD simulations will provide
a complementary perspective on simulation
work already existing in the literature.

\section{Results and Discussion}
\label{results}

\subsection{Time evolution and Temperature Dependence of Structural Parameters}
\label{sec:ave}

\begin{figure}
\includegraphics[angle=-90, scale=1]{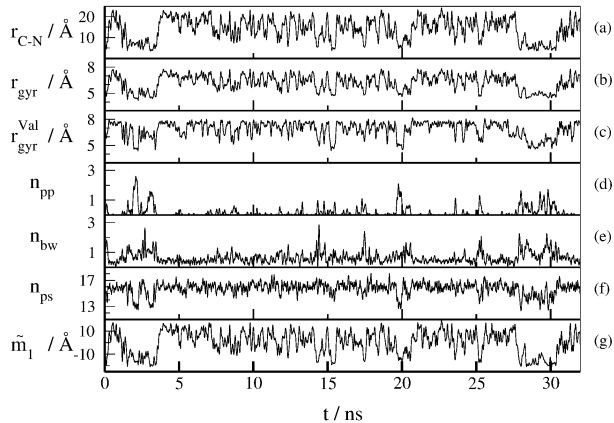}
\caption{%
Time evolution of selected structural quantities 
of GVG(VPGVG) in water at 330~K:
(a) Distance from C-- to N--terminus, $r_{\rm C-N}$, 
(b) Radius of gyration, $r_{\rm gyr}$, 
(c) Radius of gyration of the valine side groups, $r^{\rm Val}_{\rm gyr}$,
(d) Number of peptide-peptide HBs, $n_{\rm pp}$, 
(e) Number of waters bridging two amino acids of 
peptide by HBs , $n_{\rm bw}$,
(f) Number of HBs between peptide and solvating water, 
$n_{\rm ps}$,
(g) Projection of the trajectory onto the first eigenvector 
of the covariance matrix, $\tilde m_{1}$;
see text for definitions. 
Only for presentation purposes the
functions were denoised using the Savitzky-Golay
procedure~\cite{savitzky} where 
a time window of 128~ps and a polynomial of 6th degree was used,
whereas the analysis was carried out based on the original data sets.
}
\label{fig:time_struct}
\end{figure}
In order to get a first impression of the dynamics of
the solvated octapeptide GVG(VPGVG) we 
consider the time evolution
of a few selected 
quantities
describing the overall structure.
The first parameter of interest is  
the distance from the C-- to the N--terminus, 
$r_{\rm C-N}$, which is a measure of the 
extension of this short polymer chain.
A typical time evolution profile of this quantity 
is presented in 
Figure~\ref{fig:time_struct}(a) at a 
temperature of 330~K.
Judging from 
this parameter 
there are
two distinctly
different types of 
peptide 
structure: the majority of the time $r_{\rm C-N}$ 
oscillates about a distance of 
$17~\mbox{\AA}$ 
(corresponding to ``open structures'')
while
a second, less probable configuration,
is found when 
$r_{\rm C-N}\approx 5$~{\AA} (stemming from ``closed structures'').
\begin{figure}
\subfigure[ ]{\label{fig:sub:struc_open}%
\includegraphics[angle=0, scale=1]{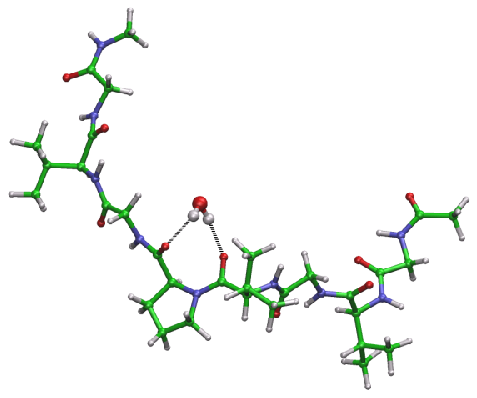}}
\subfigure[ ]{\label{fig:sub:struc_closed}%
\includegraphics[angle=0, scale=1]{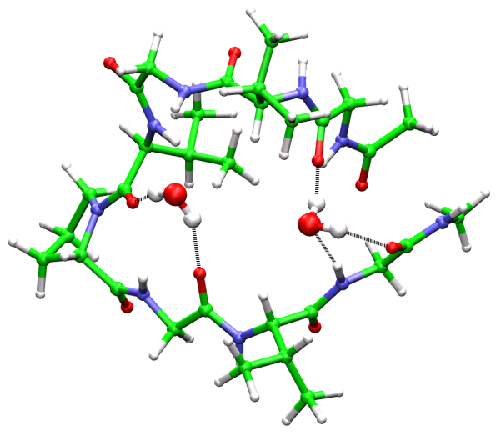}} 
\caption{%
Graphical representation of typical
protein conformations at 330~K including only bridging waters.
(a) Open state, 
(b) Closed state. 
}
\label{fig:struc}
\end{figure}
Representative configurations, 
as obtained at 330~K for each type of conformation,
are depicted in Figure~\ref{fig:struc}.
The presence of two types of peptide conformations
is also mirrored by the radius of gyration of the biopolymer,
$r_{\rm gyr}$, which is a standard measure of a polymer's overall size.
This parameter is strongly correlated with 
the end--to--end distance $r_{\rm C-N}$, see
Figure~\ref{fig:time_struct}(b),
such that for the extended structures 
$r_{\rm gyr} \approx 8$~{\AA} whereas
for the closed ones $r_{\rm gyr} \approx 4$~{\AA}.
As a simple measure of the proximity of hydrophobic side groups of the
peptide we also consider the
radius of gyration of the valine side groups only,
$r^{\rm Val}_{\rm gyr}$,
see Figure~\ref{fig:time_struct}(c).
Again, this parameter is found to be strongly, although not perfectly,
correlated with the end-to-end distance and, as expected, the radius
of gyration of the entire peptide.

The open and closed structures have distinctly
different hydrogen bond (HB) arrangements as 
indicated by 
changes in the 
numbers of HBs of different classes. 
Specifically, we consider the number of HBs 
within the peptide that directly connect
amino acid donor (NH) and acceptor (CO) groups, 
$n_{\rm pp}$, 
the number of HBs between the peptide and water
molecules in its solvation shell, 
$n_{\rm ps}$,
and the number of water molecules that bridge
different parts of the peptide chain
by simultaneously forming HBs
to two or more non-consecutive amino acid
residues, 
$n_{\rm bw}$, 
see Figure~\ref{fig:struc}. 
According to Figure~\ref{fig:time_struct}
these quantities are again found to be strongly 
correlated to $r_{\rm C-N}$.
In the closed state peptide/peptide HBs
and water bridges exhibit values of around $n_{\rm pp}\approx 2-3$ and 
$n_{\rm bw}\approx 2$, respectively, whereas peptide/solvation water 
interactions amount to 
about $n_{\rm ps}\approx 12$ HBs.
This is quite different in the extended state:
there are few peptide/peptide HBs, $n_{\rm pp}\approx 0$, 
and bridging waters, $n_{\rm bw}\approx 0$,
but about $n_{\rm ps}\approx 16$ peptide/solvation water HBs. 
These simple parameters~-- as well as others not depicted here
such as the moments of inertia~--
indicate several opening and closing events in this trajectory and thus
suggest that the present MD simulations are sufficiently able to sample the
conformational space available to the peptide to
distinguish between two very different types of structure
which are found to be in dynamic equilibrium.

A similar picture
where one observes two types of
structures, open and closed, 
with correlated $r_{\rm C-N}$, $r_{\rm gyr}$ and 
HB types is also found in simulations 
where
 the thickness of the solvating water sphere 
is increased as well as
with alternate peptide chain capping groups
at selected temperatures of 280 and 320~K 
(see Appendix~\ref{app:sys}).
This gives us confidence 
that this scenario is not an 
aberration
of our computational methodology.
Moreover,  
 in agreement with Ref.~\cite{rees98}, this latter
 finding 
suggests that the results should depend on only the 
 presence of the repeat unit VPGVG itself.
We note in passing that the time scale of the peptide backbone 
fluctuations which interconvert open and closed conformations 
is on the order of about one nanosecond 
(see Sect.~\ref{sec:pca} and in particular 
Figure~\ref{fig:sub:dynprotb}
for details) 
even for this minimalistic elastin model!
The implications of this time scale
is that for MD trajectories on the order of only a few 
nanoseconds statistical quantities,
which depend on the conformation of the peptide backbone, 
will be subject to large errors and hence should
be conservatively interpreted, 
in particular if longer chains, yielding 
even longer relaxation times, are investigated.

\begin{figure}
\subfigure[ ]{\label{fig:sub:rcn}%
\includegraphics[angle=-90, scale=0.15]{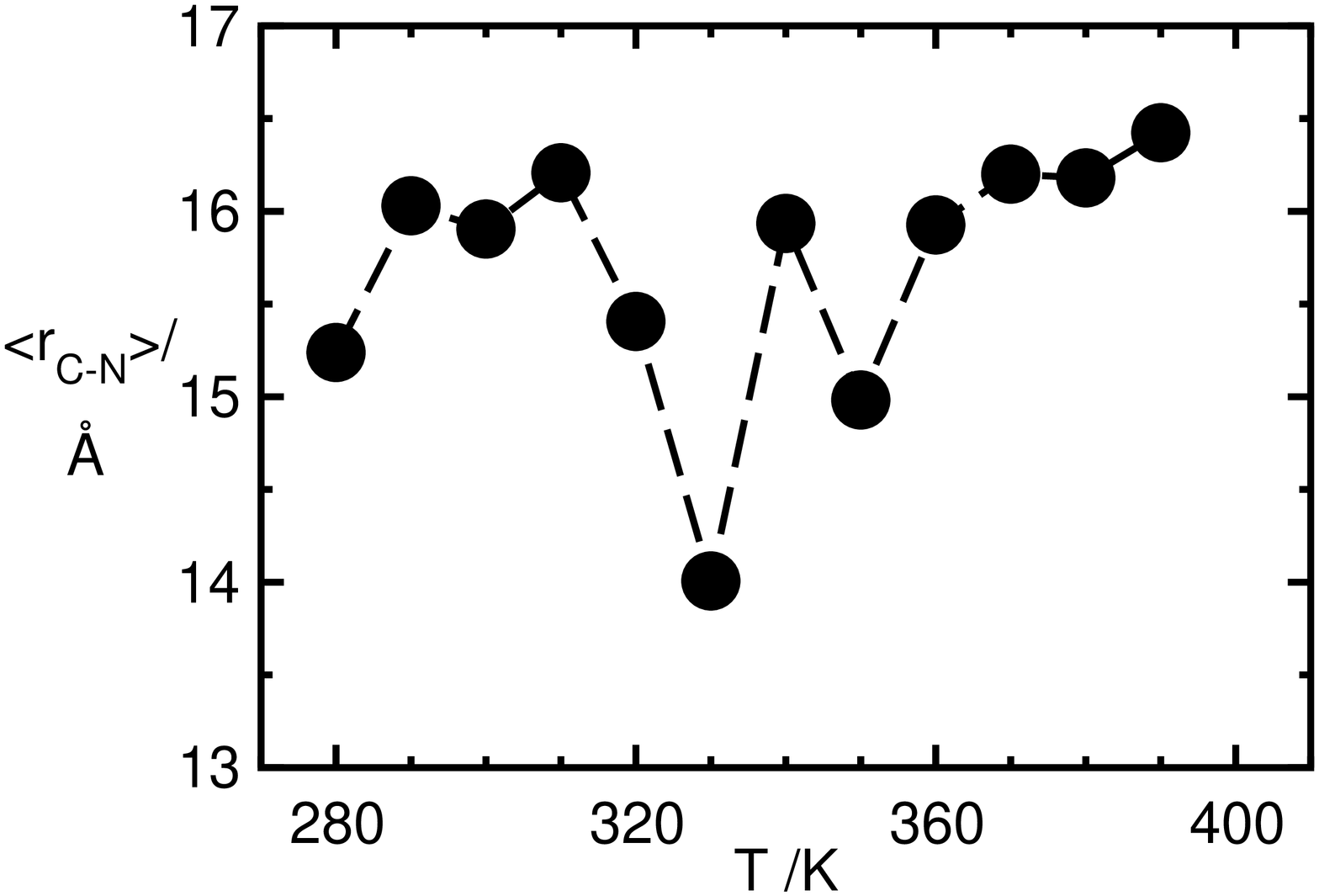}}
\subfigure[ ]{\label{fig:sub:rgy}%
\includegraphics[angle=-90, scale=0.15]{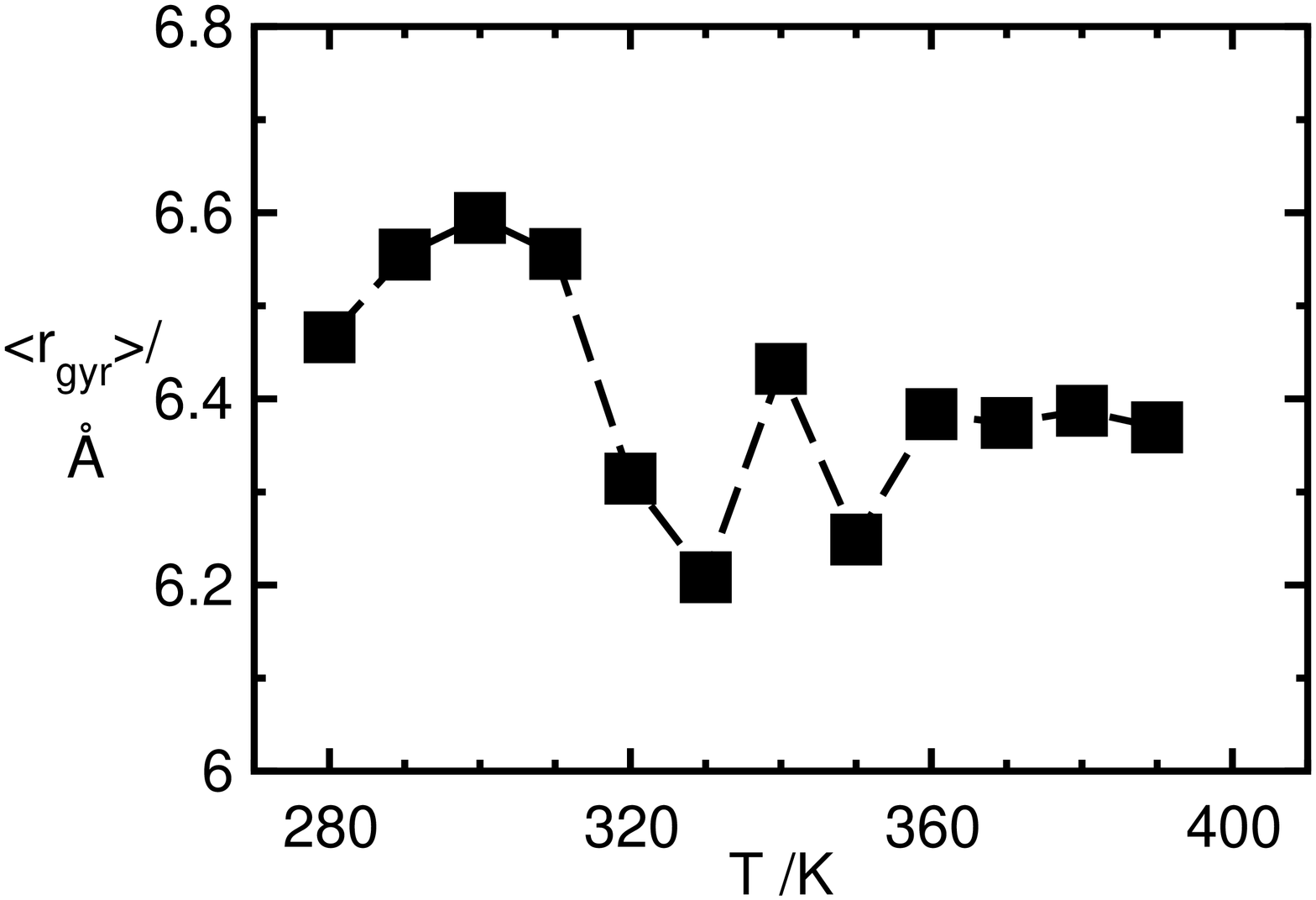}}
\subfigure[ ]{\label{fig:sub:rvalgy}%
\includegraphics[angle=-90, scale=0.15]{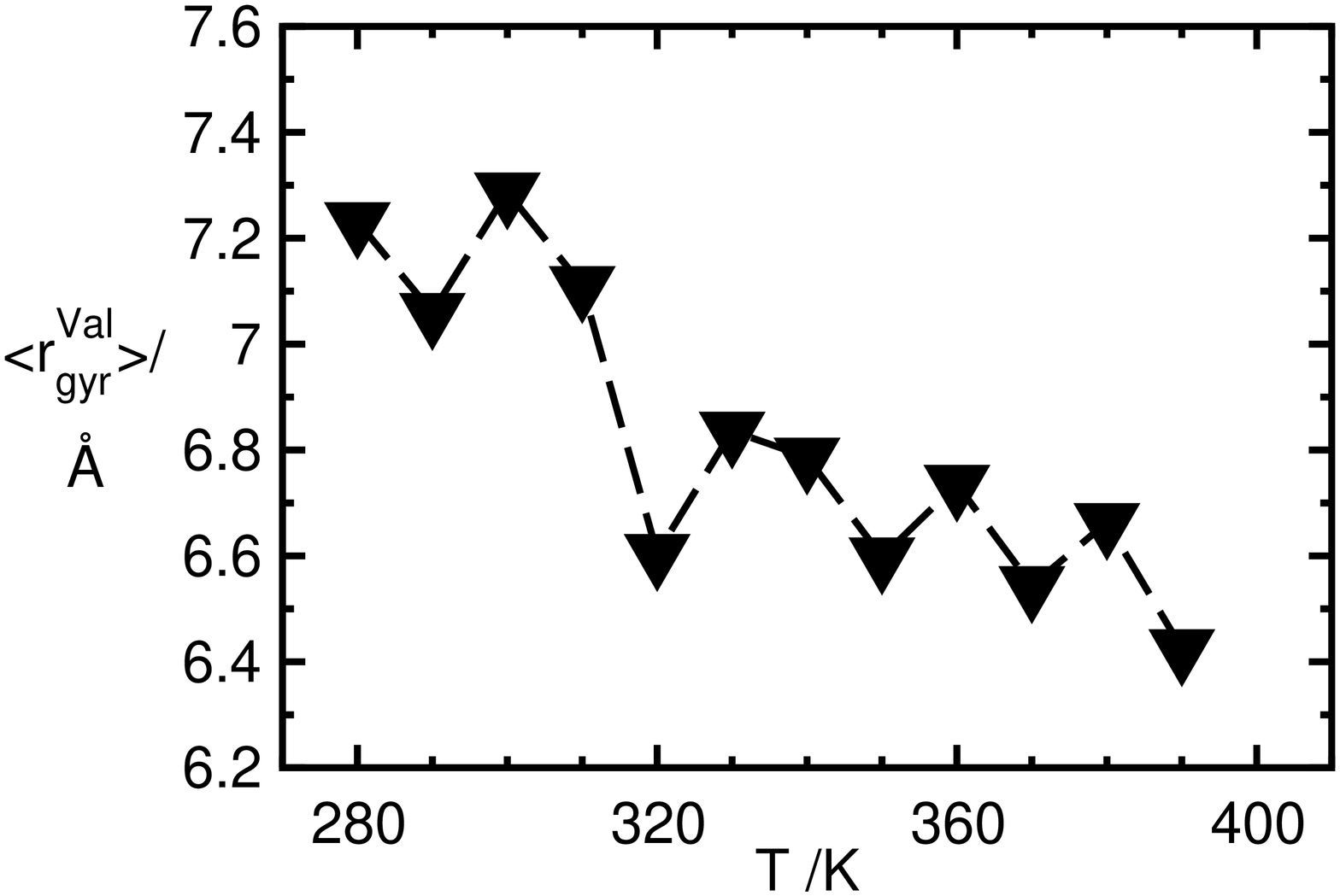}}
\subfigure[ ]{\label{fig:sub:nps}%
\includegraphics[angle=-90, scale=0.15]{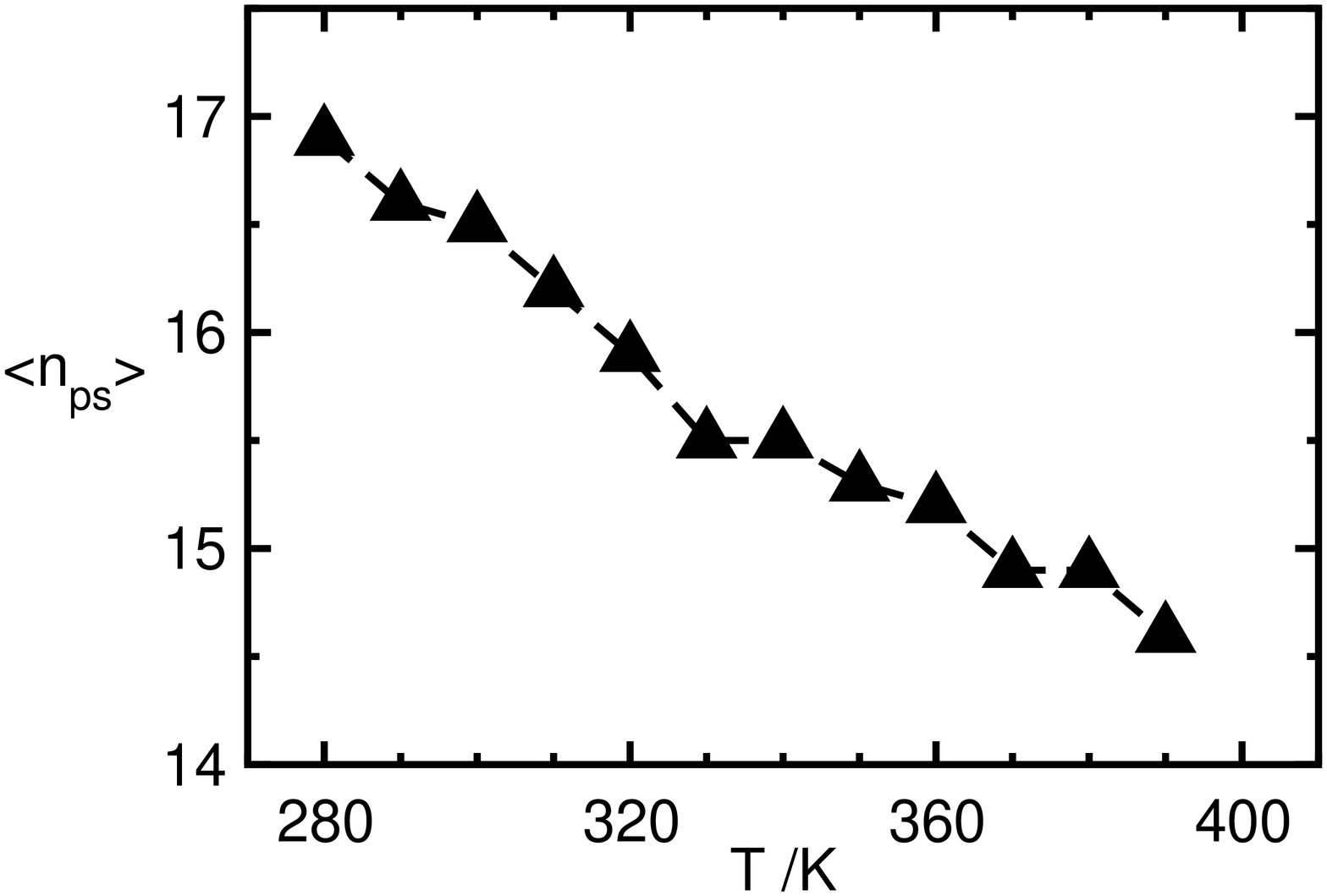}}
\subfigure[ ]{\label{fig:sub:nihb}%
\includegraphics[angle=-90, scale=0.15]{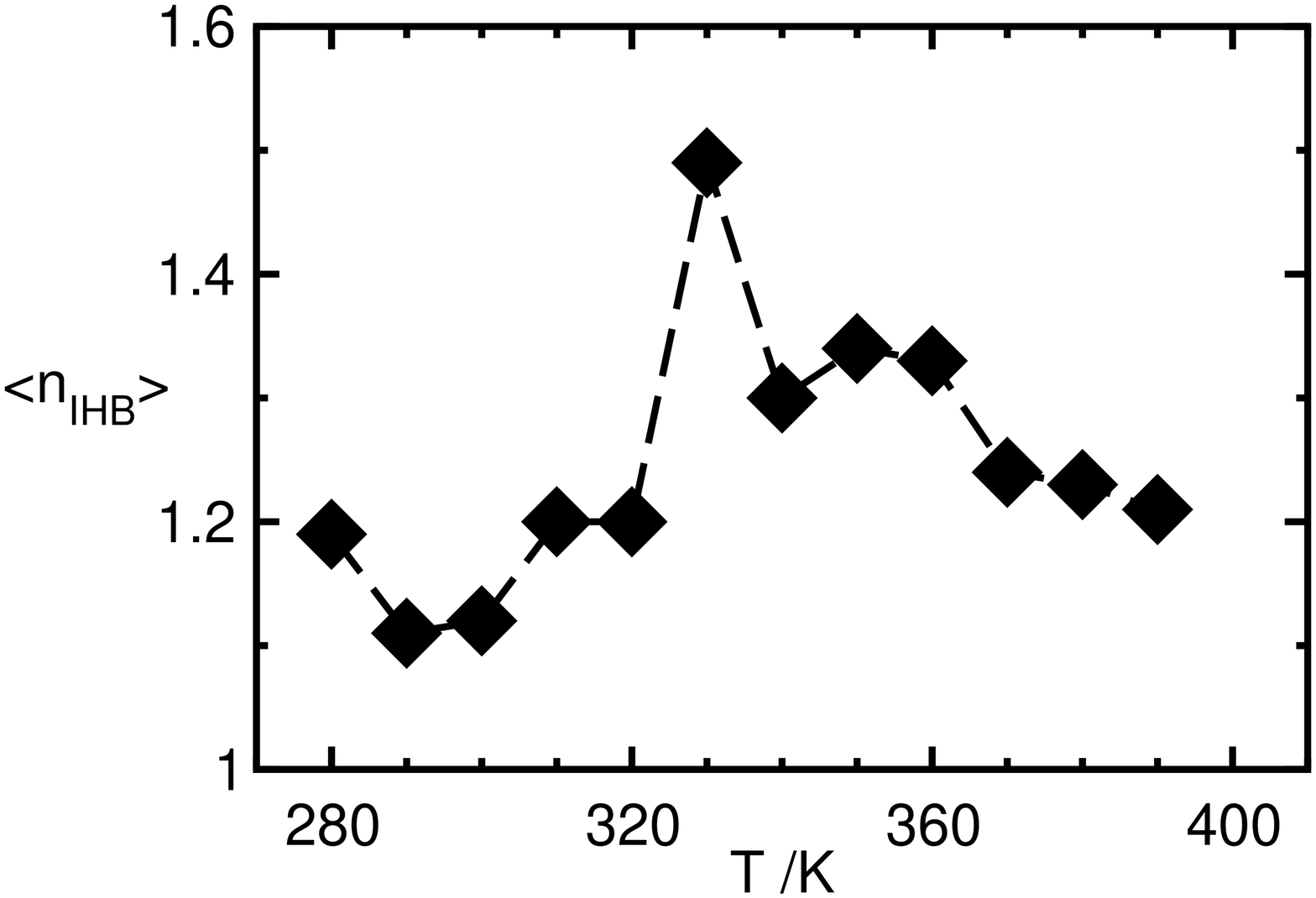}}%
\caption{%
Average quantities as a function of temperature:
(a) Distance from C-- to N--terminus, 
    $r_{\rm C-N}$ (circles),
(b) Radius of gyration, $r_{\rm gyr}$ (squares),
(c) Radius of gyration of valine side groups, $r^{\rm Val}_{\rm gyr}$
(triangles down),
(d) Number of HBs between peptide and solvating water, $n_{\rm ps}$
(triangles up),
(e) Number internal HBs, $n_{\rm IHB}$ (diamonds); see text for definitions.
Dashed lines 
are linear connections of the data to guide the eye. 
}
\label{fig:ave_struct}
\end{figure}

We now consider how this dynamic equilibrium 
behaves as a function of
temperature.
Given the above description of the change in 
HB configuration as a function of the type of structure,
it is convenient to define
 the number of internal hydrogen bonds, $n_{\rm IHB}$
as the sum of $n_{\rm pp}$ and the HB arising from the bridging
waters, $n_{\rm bw}$.
Note that both $n_{\rm pp}$ and $n_{\rm bw}$ exist only when the structure
is closed. 
An average over the entire trajectory will result in
very small numbers with large associated errors, whereas, $n_{\rm IHB}$
is a more statistically meaningful quantity 
carrying similar information.
The data from the time averages
$<r_{\rm C-N}>$, $<r_{\rm gyr}>$, $<n_{\rm IHB}>$, $<n_{\rm ps}>$, 
and $<r^{\rm Val}_{\rm gyr}>$
are plotted in Figure~\ref{fig:ave_struct}
as a function of temperature. 
As expected from the above correlations
$<r_{\rm C-N}>$,  $<r_{\rm gyr}>$, $<n_{\rm IHB}>$, 
and $<r^{\rm Val}_{\rm gyr}>$
display a similar temperature dependence
which may be classified in terms of two distinct
''temperature regimes'' below and above a 
temperature close to 330~K.
The 
average
end--to--end distance increases slightly from 280~K to 310~K,
where a significant decrease of $<r_{\rm C-N}>$ sets in
with a maximum contraction at 330~K, followed by 
a rise in  value up to 390~K.
Similarly, $<n_{\rm IHB}>$ and $<r_{\rm gyr}>$
mirror this trend. 
The average of the radius of gyration stemming
from the valine groups only, $<r^{\rm Val}_{\rm gyr}>$,
shows also an initial decrease, but at variance
with the total radius of gyration it remains 
small up to high temperatures subject to sizable fluctuations. 
Below 330~K the structures have slightly larger
$r_{\rm gyr}$ and only a few internal HBs, which is consistent with 
only a small percentage of closed structures.
The largest number of closed structures is observed  
at 330~K resulting in the smallest $<r_{\rm gyr}>$
and the largest $<n_{\rm IHB}>$.
Above this temperature,  $<r_{\rm gyr}>$ 
slowly increases in parallel to a
decrease in $<n_{\rm IHB}>$, which is consistent with a decrease in
the number of closed structures.
Note in passing that both $<n_{\rm pp}>$ and $<n_{\rm bw}>$
also follow this same trend, 
although with much 
larger numerical noise.
Overall, our picture not only is consistent with the
interpretation of an ITT occurring at around 330~K
but we also find a slow trend reversal at higher temperatures.

As a 
measure of how the peptide/water
interface behaves during these structural transitions
the total number of interfacial HBs, $<n_{\rm ps}>$, 
is considered, 
see Figure~\ref{fig:sub:nps}.
There is a continuous decrease
in $<n_{\rm ps}>$ observed upon increasing the temperature  
with a pronounced kink at around 330~K.
This hints that the peptide/water interface also 
exhibits two temperature regimes
as already observed for the structural quantities associated with the 
peptide structure.
It is noted, however, that this trend
obtained from averaging the structural quantity $n_{\rm ps}$ 
is much less pronounced 
and is examined in greater detail in Sect.~\ref{sec:hb} by 
dynamical analysis of the peptide/water interfacial HBs.

Overall, these 
averaged structural quantities 
provide a suggestive, though by no means 
comprehensive,
picture of the 
atomic level details of the peptide structural
transitions. 
The presence of two types of structures, open and closed, 
correlates extremely well with the
observation of an isodichroitic point in the CD spectra
~\cite{winter,rees98} and supports the description
of this small peptide as behaving like a two-state system.
The increase of the contribution of closed structures,
at  temperatures up to 330~K and its 
subsequent decrease at higher temperatures 
is consistent with the temperature dependence 
of more compact structures containing $\gamma$ and $\beta$ turns
as observed in the FT-IR spectra of the complementary 
experimental investigation~\cite{winter}. 
The DSC and PPC thermodynamic measurements
presented in the joint experimental study
may also be interpreted as the peptide ultimately
reverting to a structurally and thermodynamically similar state
at temperatures near the boiling point of water.
In addition, the above scenario is qualitatively
in accord with  that presented 
in Ref.~\cite{daggett01} which describes 
the ITT using a similar but much longer 
model peptide.  
Specifically this 
study also finds an increasing number
of peptide/peptide HBs, an overall average
contraction of the peptide at the ITT, 
a decreasing number of interfacial water molecules 
and an increasing number of bridging
waters 
which may be thought of as the internal waters in our small 
elastin-like analogue.
The authors of Ref.~\cite{daggett01}
attribute this shrinkage to
a ''hydrophobic collapse'' which is 
in agreement with our observed
temperature dependence of $<n_{\rm ps}>$ 
and $<r^{\rm Val}_{\rm gyr}>$.
However, the current picture suggests
the subtle but crucial difference that
there is a net exchange of peptide/peptide and bridging water 
intramolecular HBs for peptide/water interfacial HBs;
the energetic implications of the various HB types
are addressed in Sect~\ref{sec:hb}.
At this point we stress that
our finding and its interpretation  has a striking resemblance to
the recently advanced description of self--association
of methanol in water~\cite{soper} whereby
the arrangement of
hydrophilic--hydrophilic interactions are assumed
to determine the structure as opposed to entropic
hydrophobic--hydrophilic interactions.
In the following two subsections these suggestive findings, 
which are based on simple descriptors, 
will be further scrutinized by
a detailed statistical-mechanical analysis.

\subsection{Principal Component Analysis and Order Parameter}
\label{sec:pca}

To probe the above findings in a systematic but simple way one would 
ideally
wish to describe these phenomena by a single parameter, 
i.e. by an order parameter or reaction coordinate.
To explore this possibility we have employed 
what is called covariance,
essential dynamics, or
principal component analysis (PCA)~\cite{garcia,amadei93,ica}
(see Appendix~\ref{app:ana} for a short introduction).
In this approach one computes, by time averaging, $< \dots >$, 
over the entire MD
trajectory the covariance matrix
$ {\bf C}=< [{\bf x}(t) - <{\bf x}>] \cdot [{\bf x}(t) - <{\bf x}>]^T >$,
where ${\bf x(t)}$ are the Cartesian coordinates of the 
peptide atoms at time $t$ in a frame of reference where the overall 
translations and
rotations of the polymer have been subtracted.
The $3N-6$ normalized eigenvectors $\{ {\bf m}_i \}$ with nonzero
eigenvalues $\{ \lambda_i \}$  
provide a basis in which the complex 
polypeptide motion may be decomposed;
more technical details are compiled in Appendix~\ref{app:ana}.
This then provides a theoretical
framework which 
can be though of being analogous to 
harmonic normal modes that are of
ubiquitous use to explain spectroscopic observations 
of small molecules.
In practice this allows us to   
disentangle relevant 
peptide backbone folding motions
from small--amplitude vibrations via
projecting
the deviations of the MD trajectories from the average structure
onto the eigenvectors, i.e. 
${\tilde m}_i = [{\bf x}(t) - <{\bf x}>] \cdot {\bf m}_i$.
Ideally, for uncorrelated 
small--amplitude vibrations the corresponding projections
${\tilde m}_i$ will simply
fluctuate about an average value yielding  
an approximately Gaussian distribution function,
$P(\tilde m_i)$;
the width might be temperature--dependent akin to
quasi--harmonic molecular vibrations.
Strong deviations from such a unimodal behavior would single out
``interesting modes'' that can be associated to collective changes
involving many atoms.
Upon changing the temperature the various modes are
expected to mix, in particular the higher--order modes. 
Thus, a common basis $\{ {\bf m}_i \}$
has to be chosen to ensure a consistent analysis.
In the following, projections $\{ {\tilde m}_i \}$
will be performed onto the basis of 
eigenmodes of the low--temperature trajectory at 280~K.

\begin{figure}
\subfigure[ ]{\label{fig:sub:cova}%
\includegraphics[angle=-90, scale=1]{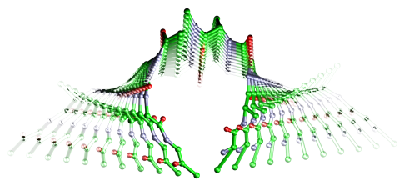}}%
\subfigure[ ]{\label{fig:sub:covb}%
\includegraphics[angle=-90, scale=1]{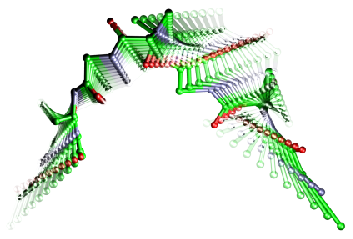}}\\
\subfigure[ ]{\label{fig:sub:covc}%
\includegraphics[angle=-90, scale=0.15]{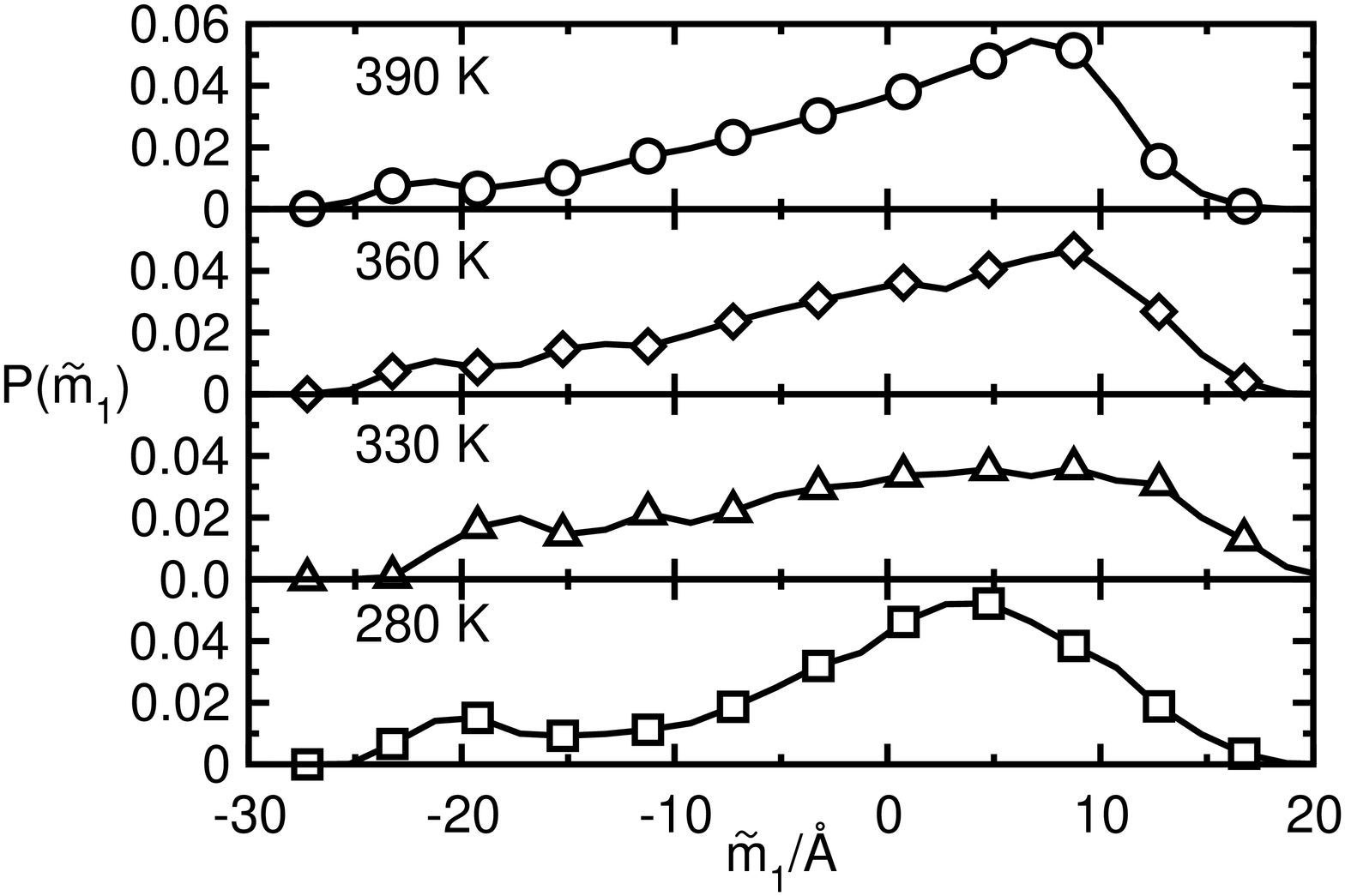}}%
\subfigure[ ]{\label{fig:sub:covd}%
\includegraphics[angle=-90, scale=0.15]{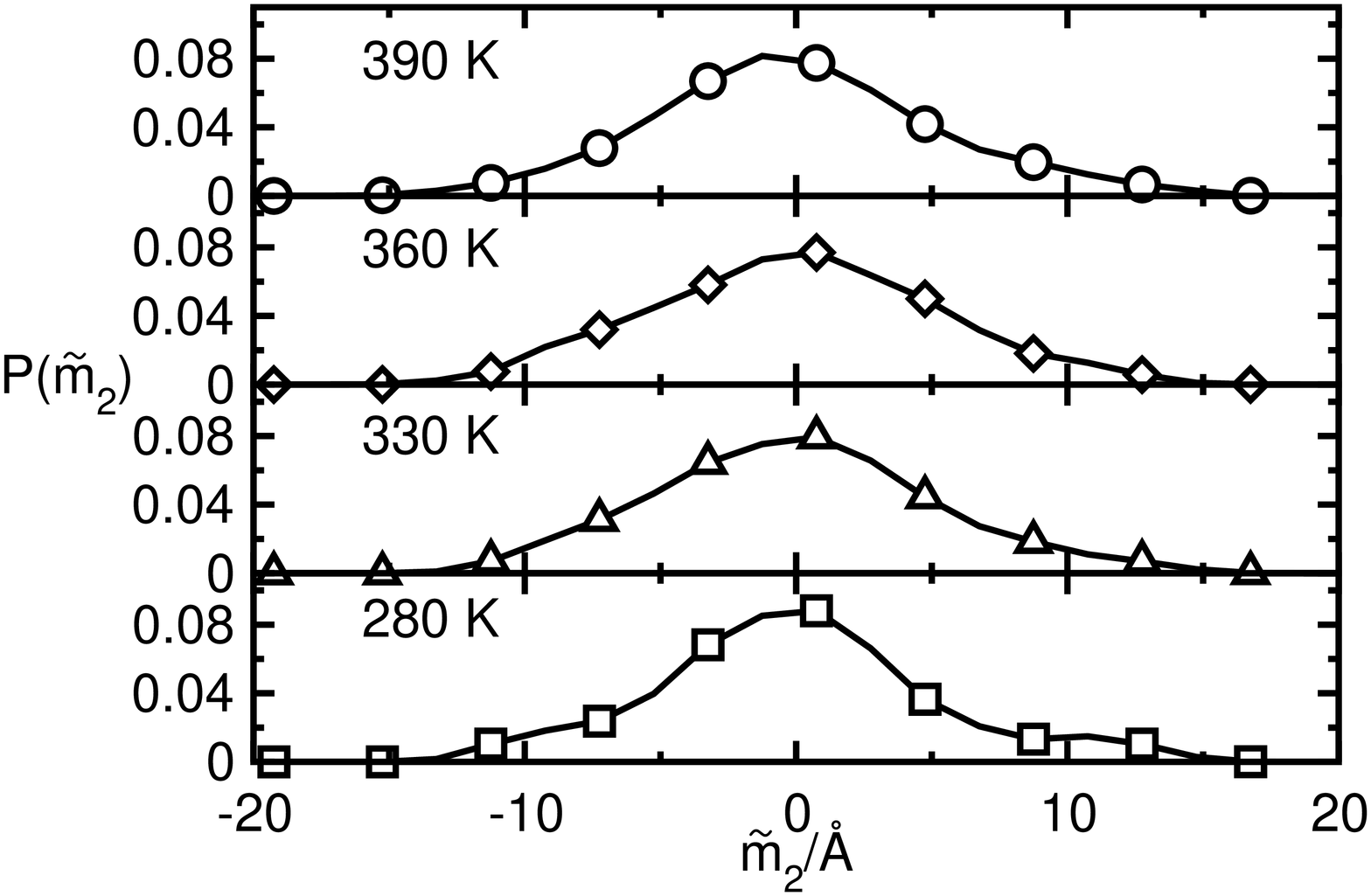}}\\
\vfill%
\subfigure[ ]{\label{fig:sub:cove}%
\includegraphics[angle=-90, scale=0.15]{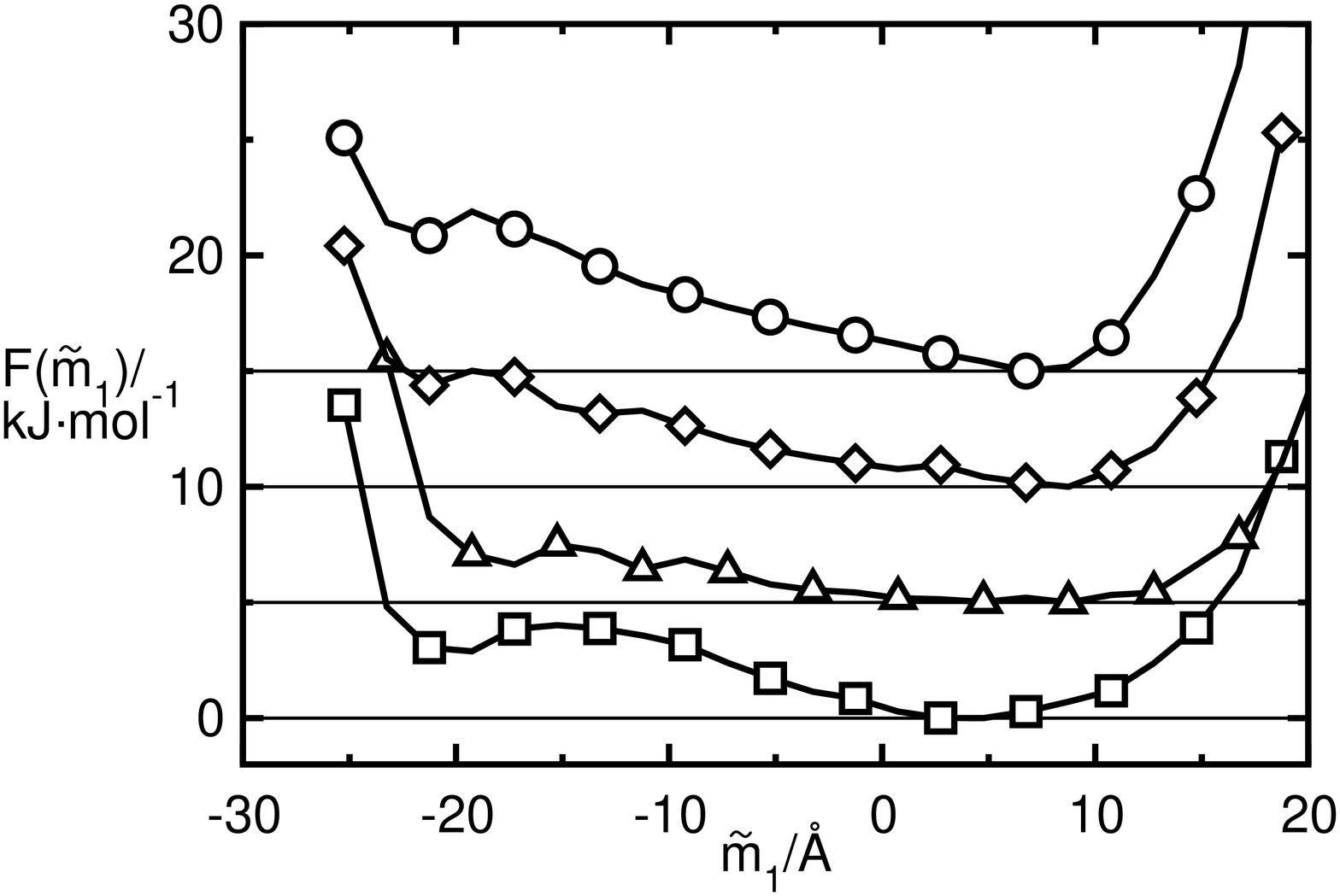}}%
\subfigure[ ]{\label{fig:sub:covf}%
\includegraphics[angle=-90, scale=0.15]{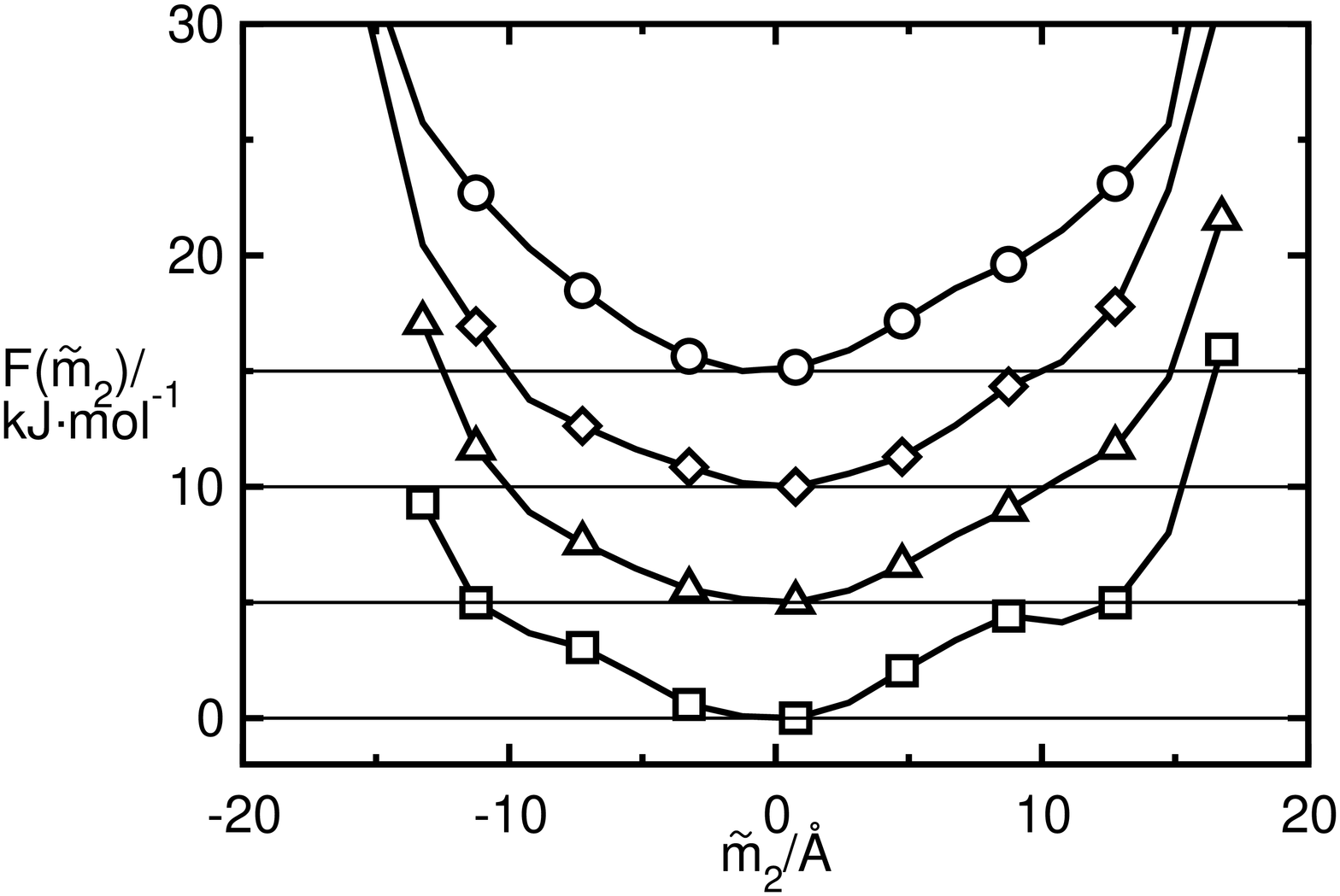}}%
\caption{%
Results of the Principal Component Analysis:
(a) 
Visualization of large amplitude opening and closing mode, {\bf m}$_1$
(b) 
Visualization of large amplitude librational mode, {\bf m}$_2$
(c) Distribution function of the projection $\tilde m_1$ at 280, 330, 360 and 390~K.
(d) Distribution function of the projection $\tilde m_2$ at 280, 330, 360 and 390~K.
(e) Relative free energy along the projection $\tilde m_1$.
(f) Relative free energy along the projection $\tilde m_2$.
In (a) and (b) an artifical trajectory using the 
full first ${\bf m}_1$, (a), and
second ${\bf m}_2$, (b), eigenvector was 
synthesized for a suitable graphical presentation.
In (e)-(f) 280 K(squares), 330 K(triangles) 380 K(diamonds)
and 390 K (circles).  
For presentation purposes the
free energy profiles for 330, 380 and 390~K were shifted
in energy by 5, 10 and 15 kJ$\cdot$mol$^{-1}$, respectively.
}
\label{fig:cov}
\end{figure}

Upon projecting the MD trajectories
onto the eigenvector with  the
largest eigenvalue $\lambda_1$ it is found that this
mode, {\bf m}$_1$, actually describes the opening and closing motion of the
octapeptide, and is present at {\em all temperatures}, 
see Figure~\ref{fig:sub:cova}
for a graphical illustration.
In Figure~\ref{fig:time_struct}(g), 
we depict this projection 
${\tilde m}_1$ as a function of time at 330~K
where it can be inferred to be perfectly
correlated with the other parameters
already discussed in the previous section.
The octapeptide thus exists in a
closed state
${\tilde m}_1 \approx -22$~{\AA} with 
end--to--end distances of about 5~{\AA}
and an extended state
${\tilde m}_1 \approx +5$~{\AA} and
end--to--end distances 
of about 17~{\AA}.
In addition, the projection process yields for ${\tilde m}_1$
a {\em bimodal} distribution function
$P(\tilde m_1)$ at low temperatures,
whereas it adopts a unimodal, but very broad and skewed shape
upon heating, see Figure~\ref{fig:sub:covc}.

The mode with the next largest eigenvalue, ${\bf m}_2$, 
can be classified as a
peptide backbone librational mode, see Figure~\ref{fig:sub:covb}
for a sketch.
During 
this motion the peptide backbone undergoes twisting 
which is actually very similar 
in spirit to that described 
for the deformation of $\beta$--turns in
Refs.~\cite{daggett01,urry-jpcB}.
However, 
in contrast to the lowest--order projection 
${\tilde m}_1$, ${\tilde m}_2$ as well as all other higher--order 
mode projections of the peptide backbone 
are characterized by fairly symmetric,
narrow and unimodal distributions $P(\tilde m_i)$ that
are essentially temperature--independent, 
see Figure~\ref{fig:sub:covd}.
This implies that~-- as far as the interpretation of the
behavior of the folding transitions are concerned~-- we
are able to neglect all backbone librational motions
and reduce our considerations down to only a 
single degree of freedom: ${\bf m}_1$.
Thus, we may confidently
employ the corresponding projection ${\tilde m}_1$
as a ``natural'' many--body collective
reaction coordinate,
or order parameter,
for investigating the
folding transition 
dynamics of GVG(VPGVG) in water.

Having obtained a proper
order parameter distribution function, $P({\tilde m}_1)$,
allows one to readily define 
an effective relative free energy profile 
(or ``potential of mean force'') according to 
$\Delta F ({\tilde m}_1)= 
- k_{\rm B}T \ln [P({\tilde m}_1)/P({\tilde m}_1^{\rm ref})]$
as a function of temperature; 
$k_{\rm B}$ is the Boltzmann constant.
The normalizing reference value $P({\tilde m}_1^{\rm ref})$ is chosen
such that the arbitrary value at the 
minimum of the free energy profile is set to zero for convenience,
see Figure~\ref{fig:sub:cove}.
At 280~K, the free energy shows two  minima, 
separated by a small barrier,
related to both open (broad global minimum at ${\tilde m}_1 \approx 5$~\AA) 
and 
closed (local minimum at ${\tilde m}_1 \approx -20$~\AA) 
structures with the latter
about 3~kJ/mol higher in energy then the former.
By 330~K the minimum associated with the closed structure,
as well as, the barrier become much lower in relative free energy
thus the entire profile of $\Delta F$ broadens out into a 
flat potential energy landscape.
Above 330~K the free energy of the closed minimum gradually 
increases 
again 
and eventually disappears by 370-390~K.
Performing an identical analysis for the
remaining modes provides approximately harmonic potentials of mean force, 
see Figure ~\ref{fig:sub:covf} for the potential derived from ${\tilde m}_2$,
in agreement with our use of only ${\tilde m}_1$ as an order parameter.
The trend in $\Delta F ({\tilde m}_1)$ reflects the presence of the  
temperature regimes observed from the average parameters, i.e. 
increased folding from 280--320~K,
the observation of 
maximum folded structures 
around 330~K,
and the unfolding at $T\geq 340$~K.
The presence of two minima in $\Delta F ({\tilde m}_1)$ 
in conjunction with the temperature-induced changes naturally
explains the existence of the isosbestic and isodichroitic points observed
in FT--IR and CD experiments~\cite{winter}, respectively, and gives us
confidence in the reliability of our simulations.

At 330~K, fluctuations and thus statistical errors are larger
then at other temperatures due to
the broader configuration space which must be sampled
possibly in conjunction with slower relaxation. 
This effect is similar to the ``slowing--down problem'' occurring 
for statistical convergence at second--order phase transitions
where all quantities, and
in particular the order parameter itself, 
undergo large fluctuations. 
However, the finite system size does not allow for a true
phase transition with a well defined 
transition temperature but rather one observes, in
agreement with spectroscopic and 
thermodynamic measurements~\cite{rees98,joint,winter}
the two  temperature regimes which are bracketed about the temperature
of maximal folding, $T\approx 330$~K.

\begin{figure}
  \subfigure[ ]
     {
            \label{fig:sub:dynprota}
            \includegraphics[angle=-90, scale=0.15]{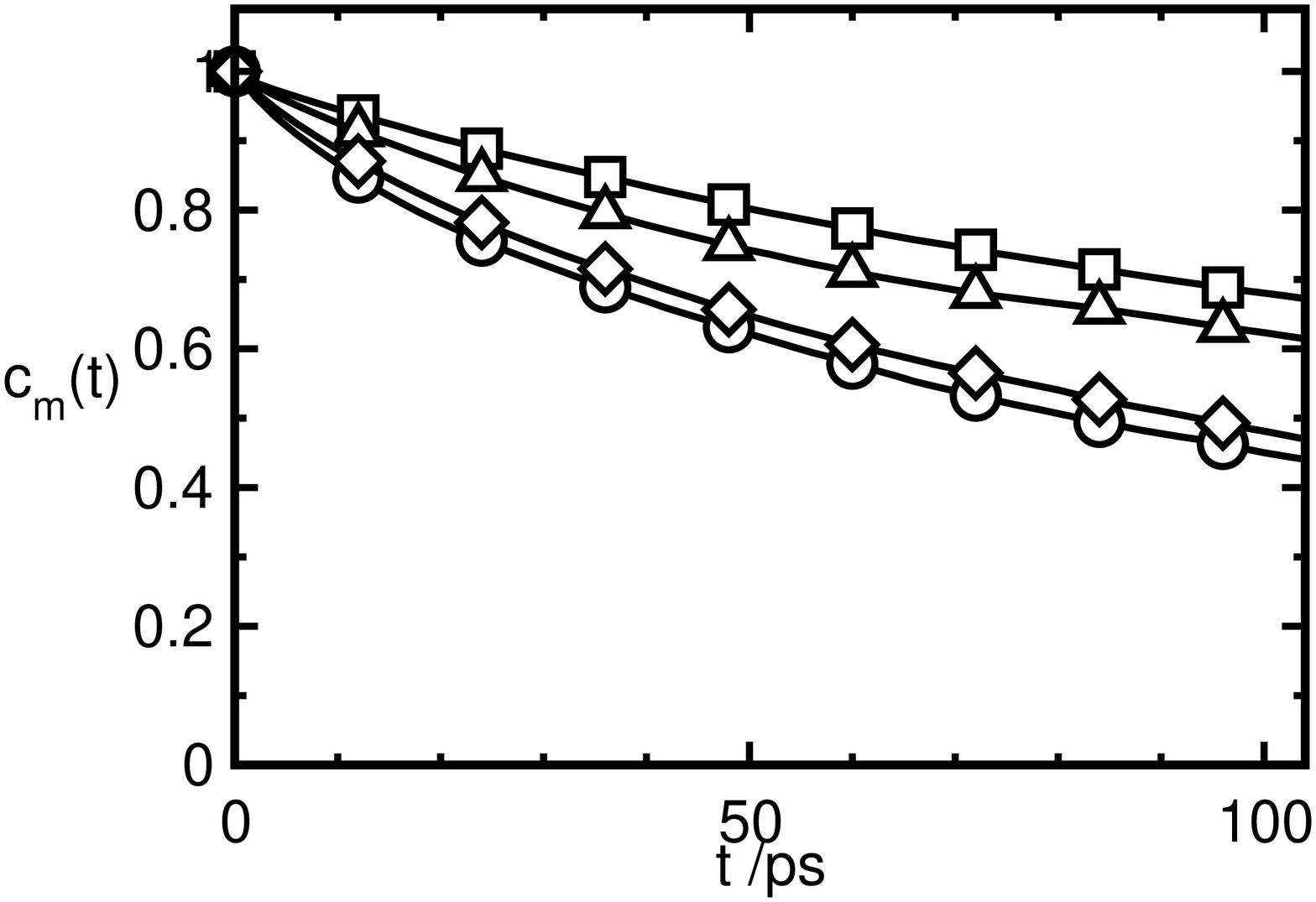}
     }
   \vspace{0.3cm}
   \subfigure[ ]
     {
            \label{fig:sub:dynprotb} 
            \includegraphics[angle=-90, scale=0.15]{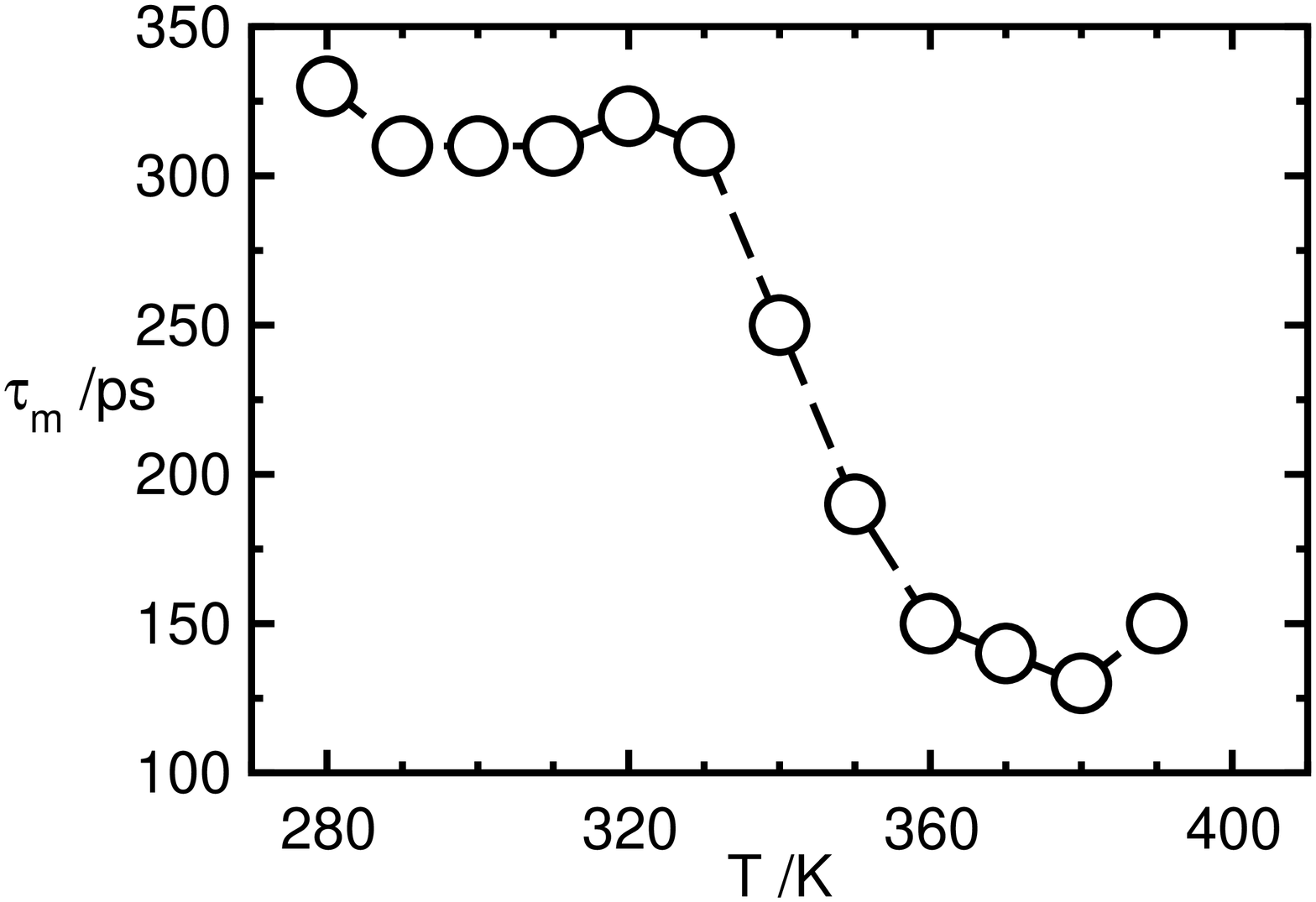}
            }
     \vspace{0.3cm}
     \subfigure[ ]
     {
            \label{fig:sub:dynprotc}
            \includegraphics[angle=-90, scale=0.15]{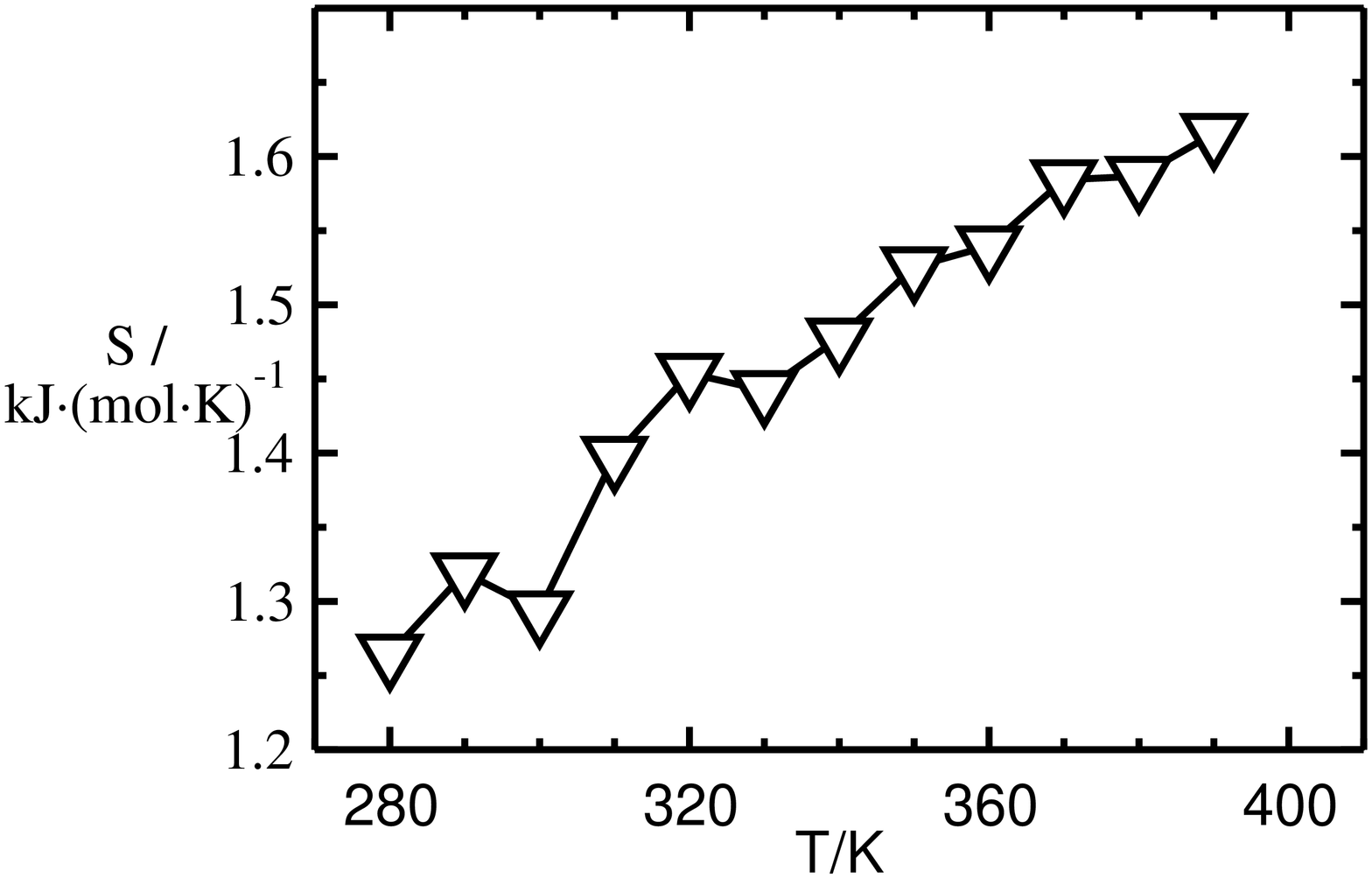}
     } 
\caption{%
Dynamics of the 
{\bf m}$_1$ eigenmode:
(a) Auto-correlation function $c_{\rm m} (t)$ 
at 280~K (squares), 330~K (triangles), 
360~K (diamonds) and 390~K (circles).
(b) Relaxation time $\tau_{\rm m}$ as a function of temperature;
here the
symbol size reflects approximate error bars.
(c) Peptide backbone quasi--harmonic entropy, $S$ (down triangles), 
as a function of temperature;
see text for definitions. 
}
\label{fig:dynprot}
\end{figure}

The dynamical motion of the peptide itself can be analyzed by 
examining the time-dependence of the reaction coordinate ${\tilde m}_1$
as described by the correlation function
$c_{\rm m} (t) = \left< {\tilde m}_1 (0) {\tilde m}_1 (t) \right> /
\left< {\tilde m}_1(0) {\tilde m}_1(0) \right>$,
see Figure~\ref{fig:sub:dynprota}.
This time auto-correlation function
may be well approximated in the short time limit, $t<100$~ps, 
by an exponential function, 
$c_{\rm m} (t) \sim \exp [ - t/\tau_{\rm m}]$,
with an associated relaxation time $\tau_{\rm m}$ 
of the peptide's opening and closing motion.
The temperature dependence of this important dynamical parameter
is depicted in Figure~\ref{fig:sub:dynprotb}.
For temperatures below 330~K,
$\tau_{\rm m} \approx 300-350$~ps 
is roughly constant within 
the accuracy of our statistics.
Above 330~K, there is a progressive speed-up
($\tau_{\rm m} \approx 130-150$ps)
which reflects the steepening of the free energy surface
in the region of closed structures, i.e. for ${\tilde m}_1 \ll 0$.
From 370--390~K, where the free energy surface contains essentially a
single minimum, ${\tilde m}_1 \approx +5$, 
with approximately constant curvature,
the relaxation time levels off to a plateau value of
about 
$\tau_{\rm m} \approx 150$~ps.
Temperature variations of $\tau_{\rm m} $
thus reflect three dynamical scenarios
 corresponding to: {\em folding}, {\em unfolding} and
{\em unfolded}, which is slightly
different from the observations 
made on the average structural quantities.
Note, however, that the last of the 
three temperature regimes represent a
limiting case of the unfolding transition
where there effectively no longer exists a mimimum for closed structures. 
These findings correlate well with larger
peptide backbone fluctuations 
 observed in NMR spectroscopy~\cite{perry02}
for water swollen elastins. 
Unfortunately, the most directly comparable NMR measurements
of  $^{13}$C relaxation times for poly(GVPGV) in solution
are only obtained for the temperatures below the ITT~\cite{Kurkova}.
None the less, the authors report a mean effective correlation time
for polymer motion on the order of about 500 ps at 298 K in 
astonishing 
agreement with the 300~ps time scale reported here
at the same temperature.
These facts suggest that temperature-dependent
measurements of NMR $^{13}$C relaxation times for this small system may
serve as an independent verification of 
our findings.

Previous MD studies on 
much
larger poly(GVPGV) chains find
qualitatively 
that the peptide becomes
``slightly more dynamic'' or  less rigid at
higher temperatures~\cite{daggett01}.
This also matches our 
quantified 
speed-up of
the 
dynamics 
of the peptide's backbone 
opening and closing motion.
Thus despite the increase in number of 
peptide/peptide HBs,
there must be an overall increase
of freedom in the motion of the peptide backbone.
The question then arises: does this extra motion lead to
favorable entropic contribution to the ITT from the peptide?
To address this, we consider the  entropy change of the
peptide backbone motion 
as estimated from a quasi-harmonic approximation
based on the principal component analysis~\cite{schlitter93,karplus01}.
Note, this analysis is based on a quasi-harmonic
approximation~\cite{schlitter93,karplus01}, whereas the 
lowest-order mode, ${\bf m}_{1}$,  
which contributes about 40\% to the overall variance of the peptide
backbone, is strongly anharmonic and even
bimodal at low temperatures. 
However, its 
contribution to the total peptide backbone entropy is
fairly small with much larger contributions arising from
the more harmonic higher-order modes.
Figure~\ref{fig:sub:dynprotc} shows the entropy of the peptide in
a united--atom representation (see Appendix~\ref{app:ana}) as a function of
temperature. 
We find that despite the volume contraction of the
peptide below 330~K the entropy is increasing with rising temperature,
i.~e. the entropy 
of the peptide itself {\em increases upon folding}.
The associated entropy change 
is also consistent in magnitude with the experimentally measured
 entropy increase
of 0.15~kJmol$^{-1}$~K$^{-1}$
and 0.11~kJmol$^{-1}$~K$^{-1}$ (at 298~K and 1~bar) 
obtained from a van't Hoff analysis
of equilibrium constants
of both CD and FT--IR data~\cite{rees98,joint,winter}.
These findings are
similar in spirit to various librational entropy
models employed to explain
viscoelastic properties of elastin and its
synthetic mimetica~\cite{urry-angew,tamburro99,urry02}.
Our analysis does not, however, eliminate the possibility that
{\em stabilizing} entropic terms
from the water are also important contributors to the folding
behavior. 
Nevertheless,
our simulations clearly demonstrate that
our elastin mimic is a very dynamical entity and that the 
thermodynamic consequences
of this motion, in particular the increasing entropy, 
are a key component in understanding its
structural transitions.

\subsection{Hydrogen Bond Dynamics and Kinetics}
\label{sec:hb}

In order to address the role of water
and in particular the influence of hydrogen bonding 
on the peptide's conformational transitions we analyze
the dynamics of the HB network.
To this end, we employ
the intermittent HB auto-correlation 
function~\cite{FHS,chandler96a,chandler96b,SNS,AC,BJB},
$c(t) = \left< h(0) h(t) \right>/\left< h \right>$.
As usual the HB population variable $h(t)$
is defined to be unity, $h(t)=1$, if a particular
HB exists at time $t$
and zero $h(t)=0$ otherwise;
the criteria to define a HB are compiled 
and discussed in Appendix~\ref{app:HBA}
and the average $\left< \dots \right>$ is taken over all HBs that 
were present at time $t=0$.
This function describes the probability that
a HB, which was intact at $t=0$, is intact at time $t$
independently of possible breakings (and reformations) in 
the interim time.
\begin{figure}
  \subfigure[ ]
     {
            \label{fig:sub:HBcfa}
         \includegraphics[angle=-90, scale=.25]{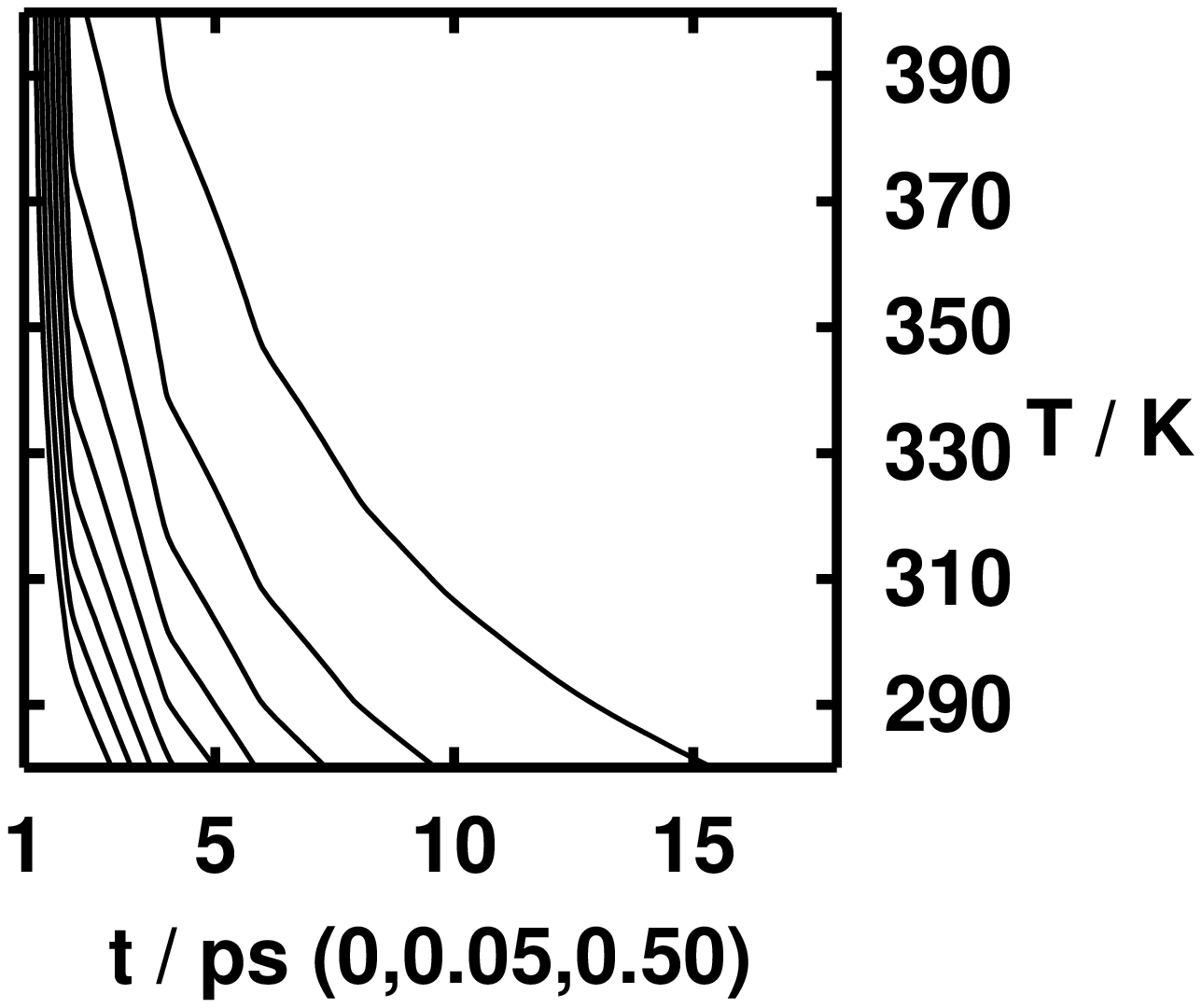}
     }
  \subfigure[ ]
     {
            \label{fig:sub:HBcfb}
          \includegraphics[angle=-90, scale=.25]{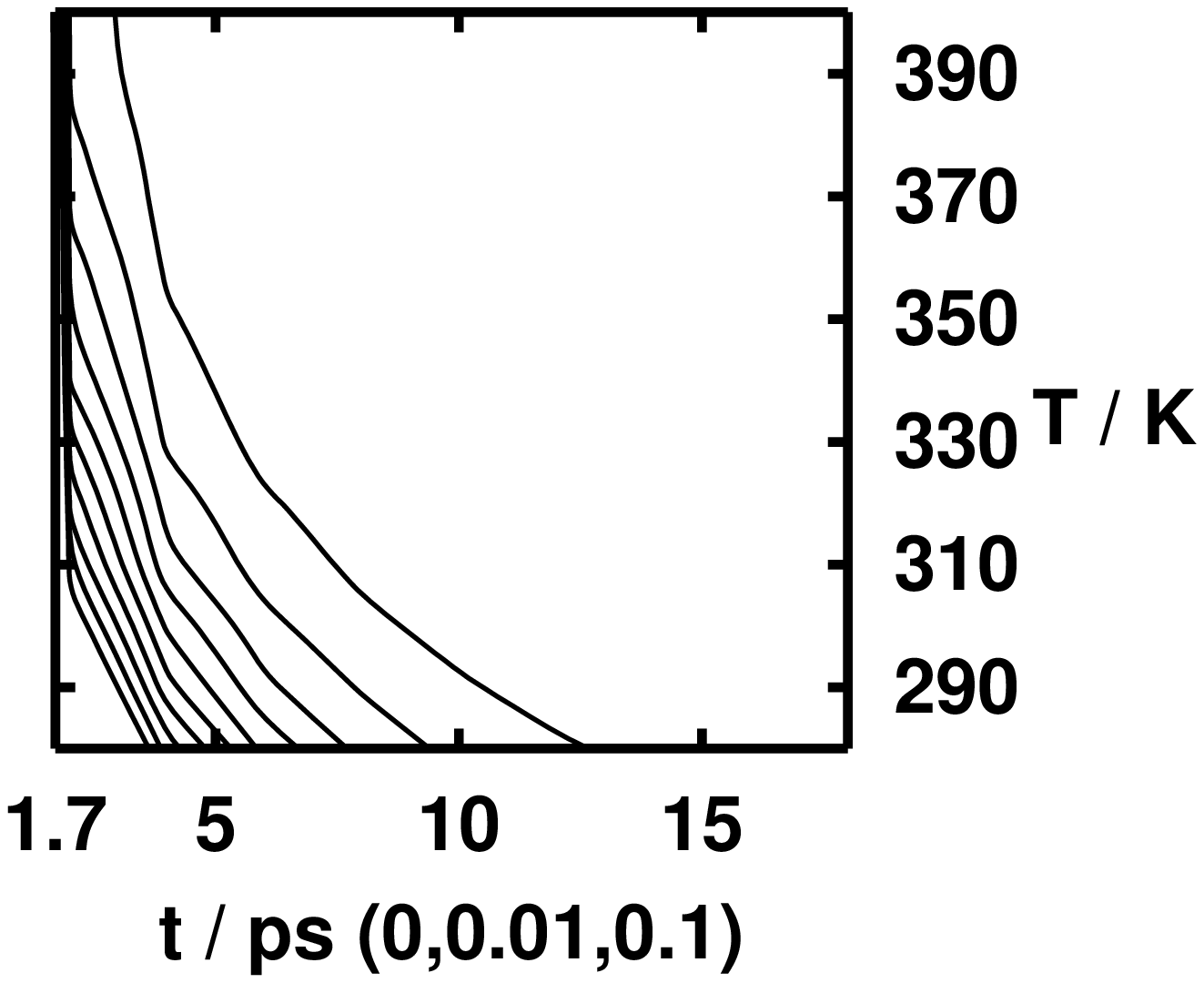}
     }
  \subfigure[ ]
     {
            \label{fig:sub:HBcfc}
          \includegraphics[angle=-90, scale=.25]{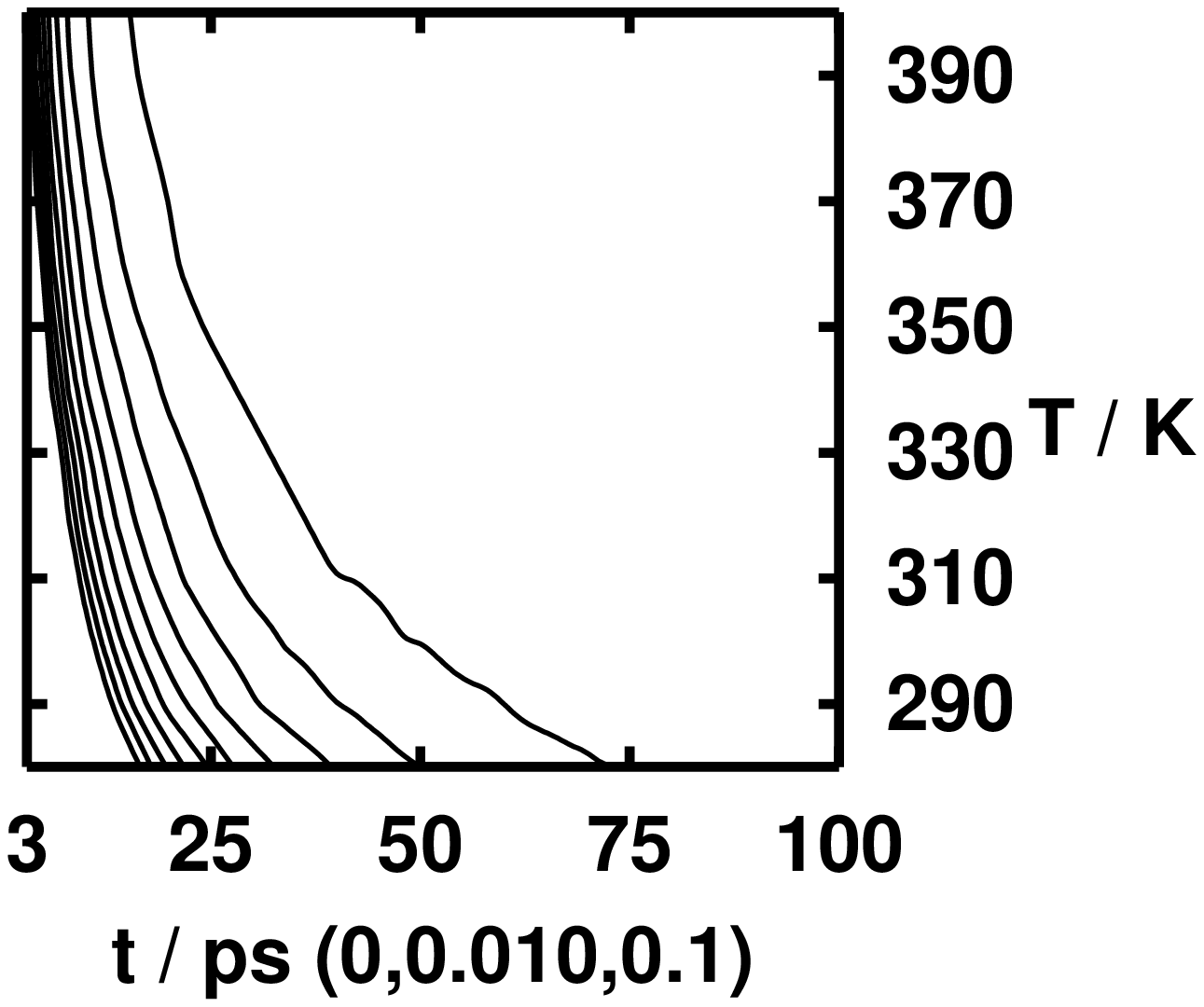}
     }
  \subfigure[ ]
     {
            \label{fig:sub:HBcfd}
            \includegraphics[angle=-90, scale=.25]{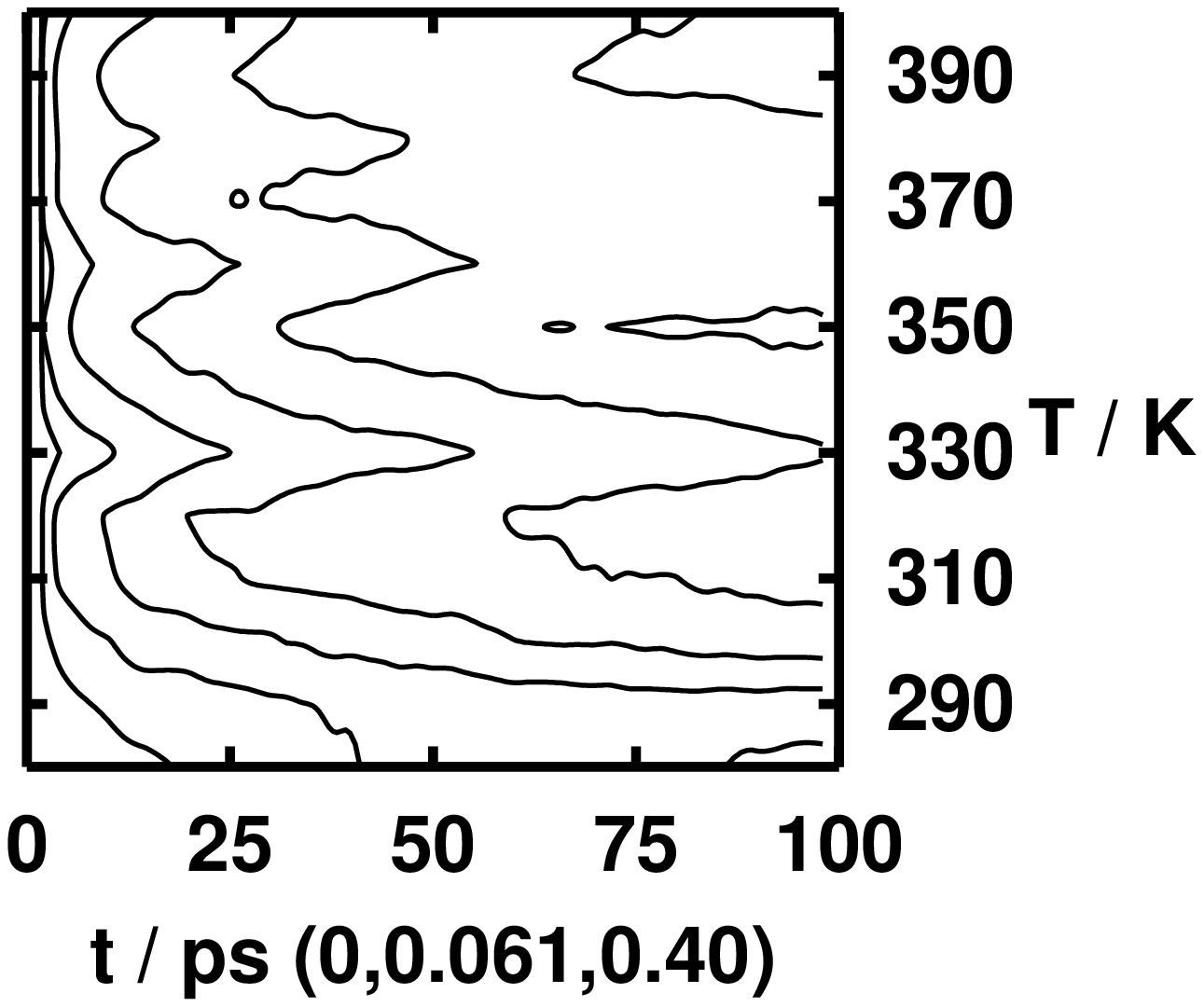}
     }
\caption{%
Temperature and time dependence of
various HB auto-correlation functions $c(t)$:
HBs between 
(a) water molecules in the bulk $c_{\rm w}$,
(b) first shell solvation water and bulk water $c_{\rm sw}$,
(c) the peptide and first shell solvation water $c_{\rm ps}$, 
and
(d) direct HB contacts of the peptide with itself $c_{\rm pp}$;
see text for definitions.
The choice of contour lines is coded in each panel as $(a,b,c)$ where
$a$ and $c$ denote the lowest and the highest contours, respectively,
and $b$ defines the relative spacing.
}
\label{fig:HBcf}
\end{figure}%
Similar to 
the average number of HBs in Sect.~\ref{sec:ave} the following
classes were introduced:
HBs between 
water molecules in the bulk $c_{\rm w}(t)$
(Figure~\ref{fig:sub:HBcfa}),
between first shell protein solvation water and bulk water $c_{\rm sw}(t)$
(Figure~\ref{fig:sub:HBcfb}),
between the peptide and first shell solvation water $c_{\rm ps}(t)$
(Figure~\ref{fig:sub:HBcfc}),
and direct HB contacts of the peptide with itself $c_{\rm pp}(t)$
(Figure~\ref{fig:sub:HBcfd}).
These functions provides us with a wealth of 
insights into the complex dynamics and the associated time scales
of the solvated peptide in relation to e.g. relaxation dynamics in bulk water. 
Comparison of $c_{\rm w}(t)$ and $c_{\rm ps}(t)$ nicely shows
the well known trend~\cite{tobias99,tobias02a,tobias02b} of
slower dynamics for interfacial peptide water HBs 
then those within bulk water:
depending on the temperature
$c_{\rm w}(t)$ and $c_{\rm ps}(t)$ decay to zero
between 10-20~ps and 20-50~ps, respectively, at temperatures below
330~K. 
At the highest 
temperatures
both functions decay to zero within 10~ps 
with the most pronounced 
speed-up of this
process occurring at the peptide/water interface.
Conversely, the function $c_{\rm sw}(t)$ 
decays to zero in only a few picoseconds at all temperatures
indicating that only 
fast HB breaking dynamics occurs for this class of bonds.
A curiosity 
is the function 
$c_{\rm pp}(t)$, which instead of decaying to zero still retains a finite
value for upwards of 100~ps or more.
This behavior arises from the fact that peptide/peptide HBs, once
broken cannot diffuse away from each other as they are 
formed involving the peptide backbone, 
i.e. they have a 
finite probability to reform even after
long elapsed times due to topology.

To extract further information on the processes
occurring on these various time scales we make use of
a more elaborate 
correlation function~\cite{chandler96a,chandler96b,luzar00,AC,SNS,BJB},
$n(t) = \left< h(0) [1-h(t)] H(t) \right>/\left< h \right>$,
This function $n(t)$
measures the {\em conditional} probability that
a HB, which was intact at $t=0$, is broken at time $t>0$
given that the donor~/ acceptor pair that established
the HB at $t=0$ is still close enough at time $t>0$ to 
potentially form again the HB.
The function $H(t)$ is unity only if the HB pair distance
is smaller than a cutoff distance of 3.6~{\AA}, which is  the second nearest 
neighbor shell of the donor~/ acceptor pair radial distribution function.
Thus, $n(t)$ allows one
to focus on the dynamics for $t\gtrsim 2$~ps where
the HB dynamics is more complex due to its coupling
to translational diffusive
motion~\cite{chandler96a,chandler96b,luzar00},
whereas the time domain for $t\lesssim 2$~ps 
is dominated
by fast librational and 
possibly
vibrational 
motion of water molecules.
\begin{figure}
 \subfigure[ ]
     {
            \label{fig:sub:HBnfa}
            \includegraphics[angle=-90, scale=.25]{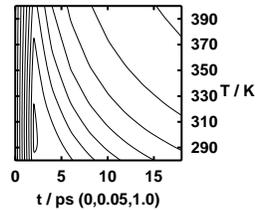}
     }
 \subfigure[ ]
     {
            \label{fig:sub:HBnfb}
            \includegraphics[angle=-90, scale=.25]{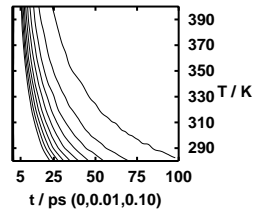}
     }
\caption{%
Temperature and time dependence of
various HB autocorrelation functions $n(t)$:
HBs between 
(a) water molecules in the bulk $n_{\rm w}$ 
and 
(b) the peptide and first shell solvation water $n_{\rm ps}$;
see text for definitions.
The choice of contour lines is coded in each panel as $(a,b,c)$ where
$a$ and $c$ denote the lowest and the highest contours, respectively,
and $b$ defines the relative spacing.
}
\label{fig:HBnf}
\end{figure}
This function plays a significant role
for two types of HBs considered in this work:
HBs between water molecules in the bulk, $n_{\rm w}(t)$
(Figure~\ref{fig:sub:HBnfa}),
and HBs between the peptide and water molecules 
in its first solvation shell, $n_{\rm ps}(t)$
(Figure~\ref{fig:sub:HBnfb}).
In both cases $n(t)$ increases from $t=0$ 
in view 
of the formation of non-hydrogen bonded pairs 
resulting from initial HB {\em breaking},
up to a maximum point 
and then decreases steadily to zero 
as the probability of  
{\em reforming} the HB decreases due to diffusion.
Comparison of Figure~\ref{fig:sub:HBnfb}(a) and (b)
shows a more rapid decay for $n_{\rm w}(t)$ than 
$n_{\rm ps}(t)$, which is due to the slower dynamics 
of HBs at the peptide/water interface.

\begin{figure}
 \subfigure[ ]
     {
            \label{fig:sub:arha}
            \includegraphics[angle=-90, scale=.15]{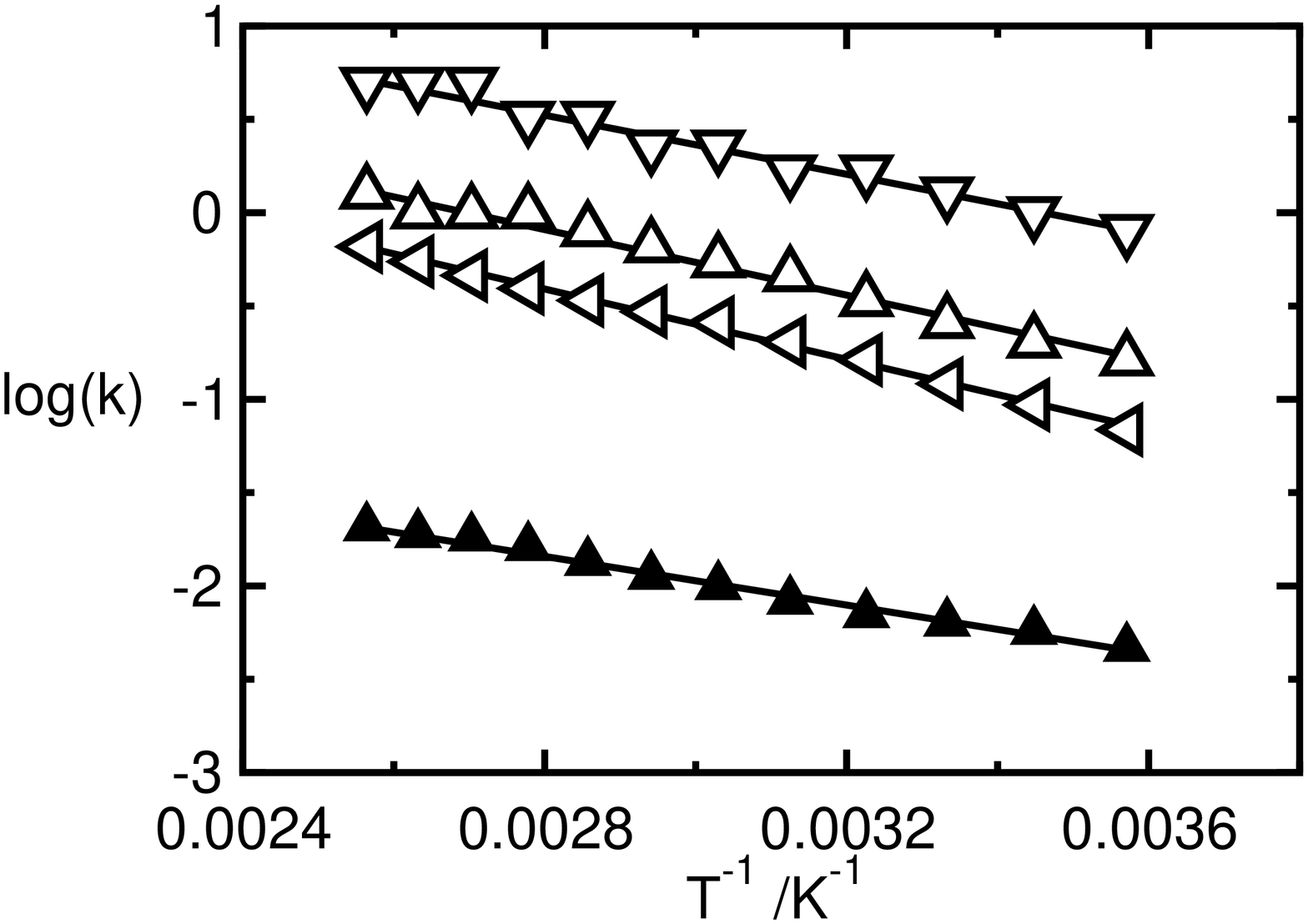}
     }
 \subfigure[ ]
     {
            \label{fig:sub:arhb}
            \includegraphics[angle=-90, scale=.15]{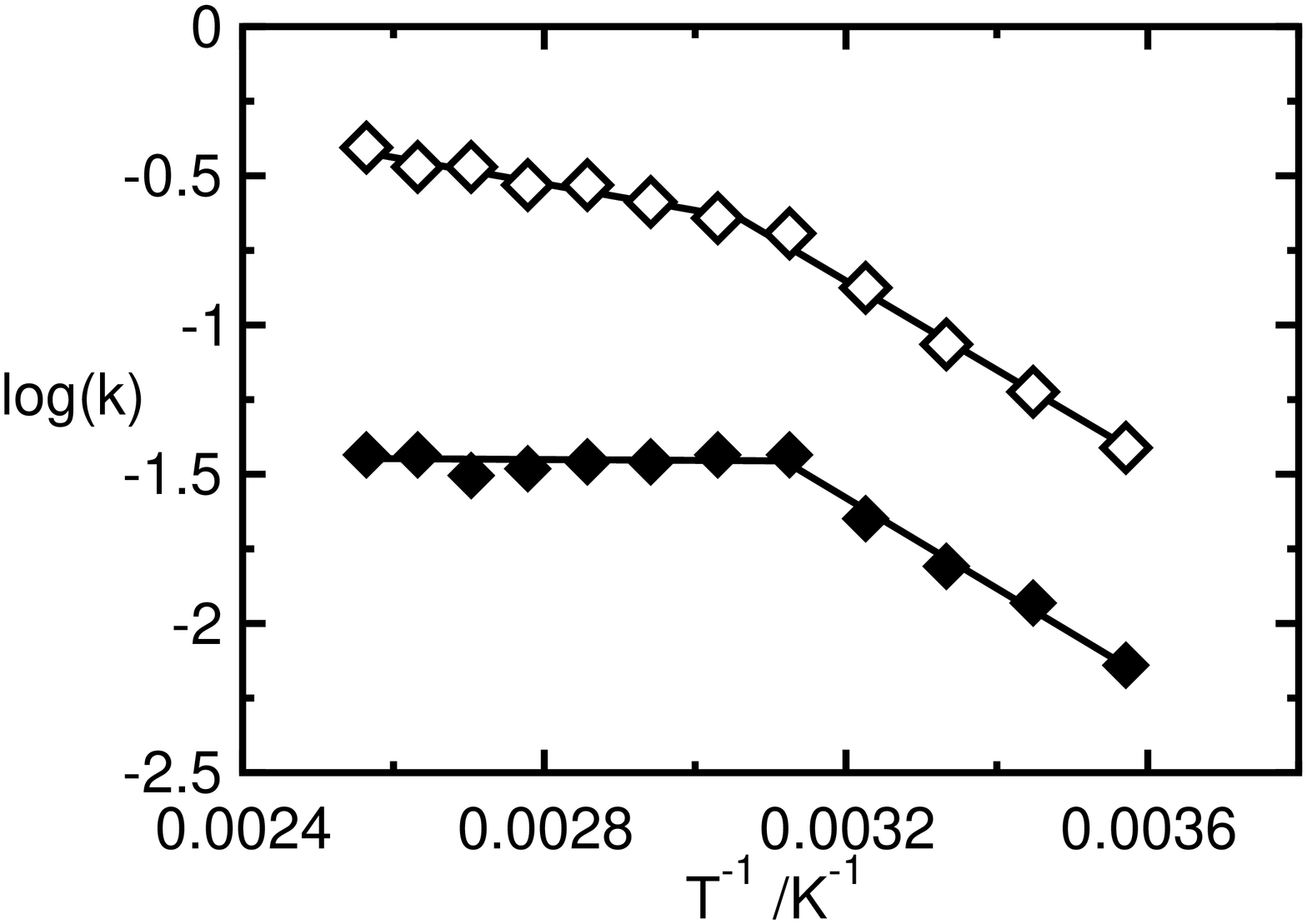}
     }
 \subfigure[ ]
     {
            \label{fig:sub:arhc}
            \includegraphics[angle=-90, scale=.15]{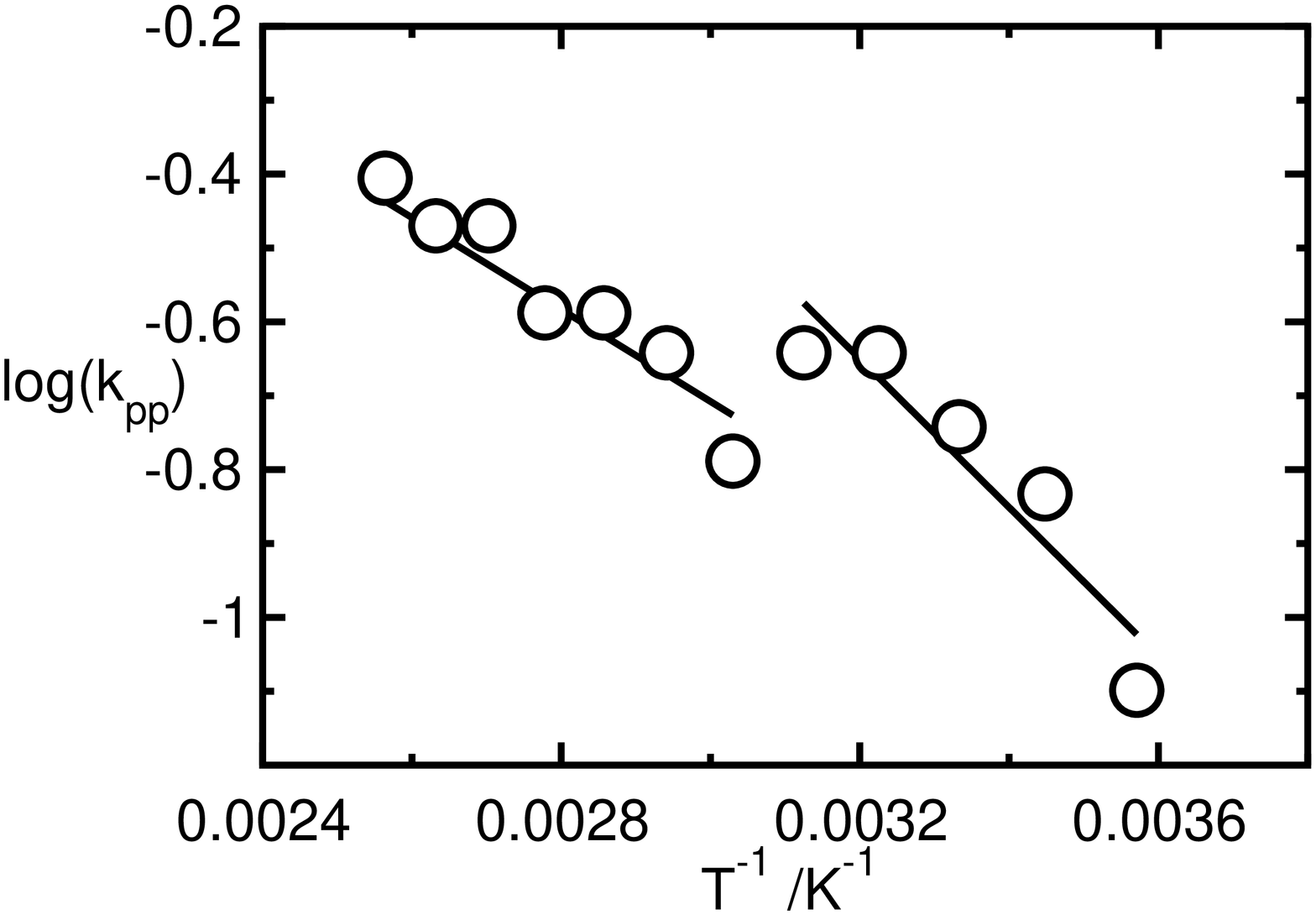}
     }
\caption{%
Arrhenius plot of rate constants for HB breaking, $k$, 
and reformation, $\tilde k$, for HBs between 
(a) 
first shell protein solvation water and 
bulk water $k_{\rm sw}$ (down triangles),
water molecules in the bulk $k_{\rm w}$ (open up--triangles)
and 
${\tilde k}_{\rm w}$ (filled up--triangles);
(b)
the peptide and first shell solvation water $k_{\rm ps}$ (open diamonds) 
and
${\tilde k}_{\rm ps}$ (filled diamonds);
(c) 
the peptide with itself $k_{\rm pp}$ (circles).
The symbol size covers the error bars and
the lines are linear (i.e. Arrhenius) fits;
see text for definitions
and Table~\ref{tab:arrh} for the resulting fit parameters.
}
\label{fig:rates}
\end{figure}

Following previous 
studies~\cite{chandler96a,chandler96b,luzar00}, 
we formulate a rate
equation describing the kinetics of HBs as a combination
of terms resulting from {\em breaking} and {\em reforming} HBs 
\begin{equation}
 -\frac{dc(t)}{dt}=k c(t) - \tilde k n(t) \ ,
\label{eq:kin}
\end{equation}
where $k$ is the rate constant of HB breaking
and $\tilde k$ is that for HB
reformation subsequent to breaking;
note that $\tau =1/k$ defines an average HB lifetime.
\begin{table}
\caption{Parameters obtained from Arrhenius analysis of
rate constants obtained for HB dynamics according to Eq.~(\ref{eq:kin}).}
\label{tab:arrh}
\begin{ruledtabular}\tiny
\begin{tabular}{lcccccc}
%\hline
HB type                    & Symbol & $T$-Range  & $A$             & $E^\ddag$                & $\tilde A$      & $\tilde E^\ddag$ \\
                           &        & K        & $\rm ps{^{-1}}$ & $\rm kJ\cdot mol{^{-1}}$ & $\rm ps{^{-1}}$ & $\rm kJ\cdot mol{^{-1}}$ \\ \hline
water(bulk)-water(bulk)    & w      & 280--390 & 10              & 7                        & 1               & 5 \\ 
water(bulk)-water(surface) & sw     & 280--390 & 15              & 7                        & $-$             & $-$ \\ 
peptide-water(surface)     & ps     & 280--320 & 60              & 12                       & 17              & 10 \\ 
peptide-water(surface)     & ps     & 320--390 & 2               & 4                        & 0.2             & $\approx 0$ \\ 
peptide-peptide            & pp     & 280--320 & 30              & 8                        & $-$             & $-$ \\ 
peptide-peptide            & pp     & 330--390 & 4               & 6                        & $-$             & $-$ \\ \hline
\end{tabular}
\end{ruledtabular}
\end{table}
For bulk water, see Figure~\ref{fig:sub:arha} and Table~\ref{tab:arrh},
both time constants $k_{\rm w}$ and $\tilde k_{\rm w}$
show a simple Arrhenius
or classical transition--state--theory behavior, i.e.
$k= A \exp[-E^\ddagger/k_{\rm B} T]$,
at all temperatures
(with $E_{\rm w}^\ddagger \approx 7$~kJ/mol and
$\tilde E_{\rm w}^\ddagger \approx 5$~kJ/mol).
Thus, both HB breaking and
reformation increase in rate {\em constantly} 
with temperature in the bulk, as expected for a simple
thermal process.
Note that the bulk water value of $k_{\rm w}$ at 300~K is
close to data from simulations of pure water at room temperature~\cite{chandler96a,chandler96b,luzar00,BJB}
whereas our value of $\tilde k_{\rm w}$ is smaller by a factor
of about one half.
Although some of this discrepancy may arise from systematic errors
as discussed in Appendix~\ref{app:HBA},
it is more likely due to the difference in the dynamical properties
of TIP3P water from SPC or SPC/E models 
(see Ref.~\cite{VanDerSpoel} and references therein)
used in previous studies~\cite{chandler96a,chandler96b,luzar00,BJB}.
In particular, 
TIP3P water has a larger diffusion coefficient~\cite{VanDerSpoel} 
which allows the water molecules to diffuse away at a different rate
after an HB breaking event.

Although the bulk water behaves in an expected
manner the same cannot be said for HBs that
connect the peptide to its surrounding water shell.
As with the average structural parameters and the peptide backbone
relaxation time, both  $k_{\rm ps}$ and $\tilde k_{\rm ps}$
show {\em two distinct} Arrhenius temperature regimes bracketed
around 330~K.
Below 320~K the activation energy $E_{\rm ps}^\ddagger \approx 12$~kJ/mol
to break an interfacial HB
is higher than that in the bulk,
whereas above 330~K it is much lower ($\approx 4$~kJ/mol).
Concurrently, HB reforming at the interface:
$\tilde E_{\rm ps}^\ddagger \approx  $11~kJ/mol
is close to that for breaking,
but at 330~K
the apparent activation energy
($\tilde E_{\rm ps}^\ddagger \approx 0$~kJ/mol)
vanishes.
In terms of rates, see Figure~\ref{fig:rates},
both processes to break and reform HBs at the peptide/water interface
effectively {\em slow down upon heating}
(w.r.t. a linear Arrhenius--extrapolation
of the low temperature behavior)
once a critical threshold, about 320--330~K, is surmounted.
Hence, the rate to make interfacial HBs
essentially reaches a plateau value above 330~K,
whereas the one to break these bonds continues to grow with temperature,
see Figure~\ref{fig:rates}.

This observation  of a profound change in the {\em dynamics} of 
the peptide/water interface 
is in qualitative accord
with thermodynamic measurements~\cite{joint,winter,urry02}.
In particular, 
the behavior of the
thermal expansion coefficient also indicates
an overall decrease in the ``strength'' of these interactions.
This is also consistent
with the proposed hydrophobic collapse which occurs
for the ITT of elastins~\cite{daggett01}.
However, the most intriguing aspect of this observation is
that the peptide/water HB dynamics,
which occurs on a time scale of 1-10~ps,
mirrors the dynamics of the peptide which occurs
on time scales that are two orders of magnitude larger.

We now investigate the dynamics of the HBs
between interfacial water and bulk water 
as well as those involving the bridging waters.
For the latter class 
a correlation function $c_{\rm bw}(t)$ is introduced based on a
HB population function $h_{\rm bw}(t)$ which is unity only if
a water molecule is {\em simultaneously} bound 
to two or more different residues of the peptide.
For the former of these two HB classes 
the second term in Eq.~(\ref{eq:kin})
may be neglected due to the relatively rapid
decay of $c_{\rm sw}(t)$ to zero.
We find that $k_{\rm sw}$ features essentially an 
Arrhenius behavior at all temperatures with a similar
value of $E^\ddag_{\rm sw} =7$~kJ/mol as bulk 
water, see Figure~\ref{fig:sub:arha} and Table~\ref{tab:arrh}.
This picture is entirely in accord with the proposition
that these surface water molecules have a 
hindered orientation
which strains the HBs that are formed with bulk water.
This does not allow for HB reformation after breaking, hence
the process of fast HB reformation does not occur for this class of HBs.
More surprising is that the correlation function of
the bridging waters, $c_{\rm bw}(t)$, essentially decays to
zero within 2~ps at {\em all} temperatures (not shown for that reason).
For this small peptide 
it may thus be concluded from the {\em dynamics} 
that, 
although water--mediated peptide/peptide
bridges (``internal water'')  do occur,
they play no key stabilizing role for the folded state.
As a corollary for the above assertion
of interchange of HBs to be at the 
heart of the folding~/ unfolding
behavior of elastin, the stability of the folded state must
then come from the 
peptide/peptide HBs.

We thus turn our focus to 
the function $c_{\rm pp}$ which describes the dynamics
of the peptide/peptide HBs.
Unfortunately, the poor statistics resulting from the few
HBs that exist for $T<320$~K, and the relatively poor convergence
for long-time decay only allow for a partial analysis. 
Nonetheless, we are able to make several important
observations.
First, the short-time regime $t<4$~ps can be used to
obtain an approximation to $k_{\rm pp}$ in a similar fashion as above,
see Figure~\ref{fig:rates}.
Below 330~K, 
$k_{\rm pp}$ may be interpreted
as having approximately Arrhenius behavior with
$ E_{\rm pp}^\ddagger \approx 8$~kJ/mol, see Figure~\ref{fig:rates}(c). 
This barrier is similar to that for HBs between
bulk water but less then those between the 
peptide and solvent waters; 
see Table~\ref{tab:arrh} for a detailed comparison.
Above 330~K, where a larger number of peptide/peptide
HBs improves the statistics of $c_{\rm pp}(t)$,
$k_{\rm pp}$ is again found to have 
an approximate Arrhenius behavior.
At these temperatures the 
peptide/peptide HBs exhibit a slightly lower activation barrier
for breaking of $ E_{\rm pp}^\ddagger \approx 6$~kJ/mol,
which is {\em larger} than that for peptide/water HBs
in the same temperature regime,
and will thus {\em stabilize the folded state}.
The increase of thermal energy which allows
peptide/water HBs to break more easily decreases the stability
of the open state but has less effect on the folded state
which is in part stabilized by peptide/peptide HBs.
Second, although the long time decay of the function
has too much numerical noise to provide a reliable fitting
it is  evident
that the probability to recover the same HB pair
decreases rapidly with increasing temperature above 330~K,
see Figure~\ref{fig:sub:HBcfd}.
This is in accord with the above observation of a 
speed-up of the peptide's backbone motion in this same
temperature regime.
Increasing fluctuations of the peptide backbone above 330~K
result in breaking peptide/peptide bonds more rapidly and decreasing
the probability with which they reform.
This effect will counter the stabilization
from peptide/peptide HBs for the folded state and
will eventually lead to the unfolding at higher temperature.

\section{Conclusions}
\label{conclusions}

The presented study is successfully able to 
reproduce the inverse temperature transition
of a minimal elastin model  
at about 40-50$^\circ$C.
Additionally, an unfolding transition is identified
at temperatures approaching the boiling point of water.
Our simulations reproduce the key observations
obtained from FT--IR, CD, DSC, and PPC experiments
in the companion study~\cite{winter} of the
same water solvated octapeptide.
Due to the small size of the system 
two broad temperature regimes are found
both in experiment and in the simulations:
the ``ITT regime'' (at about 10-50$^\circ$C)
and the ``unfolding regime'' at about $T>60^\circ$C 
where the structure has a maximum probability 
of being folded at $T\approx 50^\circ$C.
A detailed molecular picture involving a reaction coordinate and a
free energy profile 
for this process is 
presented along with a thorough 
time-correlation function analysis of the hydrogen bond dynamics
and kinetics within the peptide as well as 
at the peptide/water interface.
In particular, the two regimes are characterized by
changes in a dynamical two-state equilibrium 
between open and closed conformations
and a change of the HB dynamics involving interfacial water molecules.

In a nutshell, 
our data suggest a simple 
{\em qualitative} picture of the observed transitions.
At low temperatures, a relatively
strong peptide/water interaction stabilizes open conformations 
relative to closed ones.
The increase of thermal energy however decreases the stability
of the extended state but
has less effect on the folded conformation which is in part stabilized by
peptide/peptide HBs.
Thus, there is a shift in equilibrium and one observes an
increase in folded structures and hence the ITT.
A second important contributing factor is the
increase of peptide backbone fluctuations above 
the ITT and the resulting
entropic stability. 
However, the large librational 
amplitude motion of the peptide backbone
can provide sufficient thermal excitation to
break these peptide/peptide HBs at
higher temperatures
which ultimately leads to a second, 
{\em unfolding}, transition.

The current work, based upon
a temperature scan using 32~ns simulations of an octapeptide,
gives further support to the
conclusions of previous 
studies employing much shorter MD simulations on longer
poly(VPGVG) peptides~\cite{daggett01,daggett02,daggett02b,daggett02c}.
As in these studies we also observe a decrease in
the peptide/water interactions around the ITT and a speed-up of
the peptide backbone dynamics above this transition.
Although our longer simulation times
do {\em qualitatively}
agree with the findings of this earlier work the present study
is able to provide a more  {\em quantitative} analysis
of the {\em dynamical processes} associated with the peptide
structural transitions.
Novel findings related to the  inter-relation of
of peptide/water HB dynamics and
peptide librational entropy, which are 
complementary aspects of the ITT and unfolding transitions 
as opposed to being alternate models,
are only possible due to these quite
long MD trajectories. 
Thus, the current work emphasizes the importance
of detailed statistical-mechanical analysis in 
constructing atomic level pictures of complex
biomolecular processes.

The limitation of the current approach however, is that
it is not at present clear how all of the factors,
which have been isolated as key ingredients 
in the peptide structural transitions, will apply
to large polypeptides.
Although it is highly likely that decrease in peptide/water
HB interactions and peptide librational entropy 
will remain important ingredients 
the relative weights of the two 
factors in larger polymers is unknown.
Most notably,the energetic/entropic contribution
of bridging waters in larger polymers where the 
folded peptide may be truly said to have an
``inside'' is currently an open issue.
To address these questions, 
future combined experimental and theoretical work on 
GVG(VPGVG)$_n$ with $n=2$ and $3$ including mutant structures as well
as tropoelastins is currently underway.

\acknowledgments
\label{thanks}

This work has benefitted immensely from
insightful and enjoyable interactions with:
Amalendu Chandra,
Alfons Geiger,
Helmut Grubm\"uller,
Matthias M\"uller,
Chiara Nicolini, 
Nikolaj Otte,
Hermann Weing\"artner,
Roland Winter,
and 
Xiao--Ying Yu.
The simulations were carried out at {\sc Bovilab@RUB} (Bochum)
and we thank DFG (FOR~436) and FCI for partial financial support.

\appendix
\section{Theoretical Methods}
\label{comp}

\subsection{System}
\label{app:sys}

The system  consists
of an eight amino acid
oligopeptide of sequence
GVG(VPGVG)
capped by methylamine \mbox{(--NH-CH$_3$)}
and acetyl \mbox{(--CO-CH$_3$)} groups
at its C-- and N--termini, respectively.
In accordance with the experimental finding~\cite{rees98}
that VPGVG is the minimal repeat unit necessary for
observing the ITT irrespective of the end groups,
the caps are not expected to change the basic scenario
as found in 
our complementary
experimental study on the zwitter ionic species~\cite{joint,winter};
further checks are discussed below.
The octapeptide is solvated with 
2127 waters and its center of mass is fixed to the
center of a 50~{\AA} diameter
spherical droplet surrounded by
stochastic boundary conditions~\cite{stochastic}.
This sphere diameter allows us to maintain 
at least three to four 
molecules between the peptide and
the stochastic boundary even for the configurations 
where elastin is at its maximum
elongation of about 27~\AA.
We employ the
all atom
{\sc Charmm} force field~\cite{charmm22}
for the peptide 
and the {\sc TIP3P}~\cite{tip3p} water potential.

\subsection{Simulations}
\label{app:sim}

Within the  EGO molecular dynamics package~\cite{ego1,ego2}
a weak Berendsen thermostatting scheme (coupling time constant of
100~fs)
was used to control the temperature 
using a 1~fs MD time step and
the {\sc Shake} algorithm for 
fixing the 
bonds between hydrogen and heavy atoms.
This conservative choice of parameters for 
a careful 
integration of the
equations of motion has allowed us to perform
stable 
simulations which included 
equilibration times of 2--5~ns (depending on temperature)
prior to obtaining 
32~ns trajectories at 12 temperatures between 280 and 390~K.
An additional run was performed at 400~K (not shown)
in order to 
stretch the temperature range as much as possible beyond 
the boiling point of water.
For many
properties, such as HB breaking and reformation rate constants,
the same trend as obtained from ''extrapolation'' of the
data from 340 to 390~K was observed.
However, in particular 
average structural parameters showed deviations
from this trend which we interpret tentatively
as failures of the simulation approach.
Note that 400~K exceeds by about 100~K the temperature
range for which the employed force fields for both
water and the peptide are optimized, 
i.e. ambient conditions.
As a convergence check an additional 8~ns run 
was obtained at 280~K in order to test the reliability of 
both average and dynamical properties of the protein;
the results at 280~K are quoted for the 32~ns MD run for
statistical 
consistency. 
Internal pressures
inside the droplet were found to be
less than about 
0.1--0.2~kbar depending on the temperature, which is well
below the pressure of the order of many kbars that is 
required to suppress  
protein folding entirely~\cite{PTdiagram,winter}.

In order to cross--check the possible influence of the
capping groups 
we have computed two 25~ns trajectories at 
280 and 320~K in larger droplets
(60~{\AA} diameter with 3700 water molecules) for both the
capped and zwitterionic peptides.
We obtained qualitatively identical results to those presented here
indicating that the ITT is not significantly 
affected by the presence of the caps, 
which is in line with experimental findings~\cite{rees98}, 
nor by the spherical boundary
conditions for the chosen size of the solvating droplet.

\subsection{Principle component analysis}
\label{app:ana}

Principal component analysis (PCA) is a standard projection method in
statistical data analysis, feature extraction, and data compression~\cite{ica}.
Given a set of multivariate data, the purpose is to find a smaller 
set of variables with less redundancy reproducing the 
original data as well as possible. 
The redundancy is measured by 
correlations between the data elements.
       
The starting point in PCA is a random vector 
${\mathbf x}$ with $n$ elements and a sample 
{\bf x}$^{(1)}$, $\dots$, {\bf x}$^{(L)}$ from this vector.
No assumptions on a generative model or the probability 
density of the vectors are made
as long as the first- and second-order statistics 
can be estimated from the sample.
      
First of all the vector $\mathbf{x}$ is centered
       \begin{equation}
         \mathbf{x}\leftarrow \mathbf{x}-<\mathbf{x}> 
      \label{center}
       \end{equation}
and the mean $<\mathbf{x}>$ is estimated from the available sample. 
Geometrically this transformation corresponds to a translation of the entire sample
so that the mean is now at the origin.
Next consider a linear combination
       \begin{equation}
        \tilde m_1=\sum_{k=1}^{n}w_{1_{k}}x_{k}=\mathbf{w}_{1}^{T}\mathbf{x}\enspace ,
         \label{lincomb}
       \end{equation}
where the index goes over all elements of {\bf x}. The weights $ w_{1_{1}},
$\dots$, w_{1_{k}}$ are real numbers, elements of an $n$-dimensional
vector $\mathbf{w}_{1}$ and $\mathbf{w}_{1}^{T}$ denotes the
transpose  of $\mathbf{w}_{1}$. The factor $\tilde m_{1}$ is called the first 
principal component of $\mathbf x$, if the variance of $\tilde m_{1}$
is maximally large. 
Therefore one has to maximize the second moment
       \begin{equation}
         <\tilde m_{1}^{2}>=<(\mathbf{w}_{1}^{T}\mathbf{x})^{2}>
         =\mathbf{w}_{1}^{T}<\mathbf{x}\mathbf{x}^{T}>\mathbf{w_{1}}
         =\mathbf{w}_{1}^{T}\mathbf{C_{x}}\mathbf{w}_{1}
       \end{equation}
subject to the normalization condition
       \begin{equation}
         |\mathbf{w}_{1}|=(\mathbf{w}_{1}^{T}\mathbf{w}_{1})^{1/2}=1\enspace ,
       \end{equation}
       where $|\cdots|$ denotes the euclidian norm 
       \begin{equation}
         |\mathbf{w}_{1}|=(\mathbf{w}_{1}^{T}\mathbf{w}_{1})^{1/2}
       \end{equation}
       of $\mathbf{w}_{1}$
       and $\mathbf{C_{x}}$
       is the $n\times n$  covariance matrix of $\mathbf{x}$.
       The maximization problem can be rewritten
       \begin{equation}
         (\mathbf{C_{x}}-\lambda\mathbf{I})\mathbf{w}_{1}=\mathbf{0}
       \end{equation}
       or
       \begin{equation}
         \det(\mathbf{C_{x}}-\lambda\mathbf{I})=\mathbf{0}
       \end{equation}
       which means that $\lambda$ is an eigenvalue of $\mathbf{C_{x}}$
       and that $\mathbf{w}_{1}$ is the corresponding eigenvector 
       \begin{equation}
         \mathbf{w}_{1}=\mathbf{m}_{1}\enspace .
       \end{equation}
Therefore the first principal component is a projection of the
vector $\mathbf{x}$ onto the unit length eigenvector of it's covariance
matrix having the largest variance, i.e. eigenvalue.

This solution can be generalized to $n$ principal components
where the $i$-th principal component is
       \begin{equation}
          \tilde m_{i}=\sum_{k=1}^{n}m_{i_{k}}x_{k}=\mathbf{m}_{i}^{T}\mathbf{x}
       \end{equation}
with $1\leq i\leq n$ and the eigenvectors are ordered in
descending order of the magnitude of their eigenvalues
       $\lambda_{1}>\lambda_{2}>\cdots >\lambda_{n}$.
Thus the original data set can be fully reproduced by the expansion
       \begin{equation}
       \sum_{i=1}^{n} \tilde m_{i}\mathbf{m}_i=\sum_{i=1}^{n}\mathbf{m}_{i}^{T}\mathbf{x}\enspace \mathbf{m}_{i}\enspace .
       \end{equation}
 The hope is now that by considering just the first few
principal components most of the data can be represented as
       \begin{equation}
\mathbf{x}=\sum_{i=1}^{n}\tilde m_{i}\mathbf{m}_i\approx\sum_{i=1}^{j<n}\tilde
       m_{i}\mathbf{m}_i=\mathbf{x}_{\rm approx} \qquad \textrm{i.e. } \mathbf{x}_{\rm approx}\approx\mathbf{x} \enspace .
       \end{equation}
       
In the case of analyzing MD trajectories the variable 
$\mathbf{x}$ can be an $3N$-dimensional vector consisting of
the cartesian coordinates of the $N$ atoms under consideration in the
form
       \begin{equation}
         \mathbf{x}=(x_{1}, y_{1}, z_{1}, x_{2}, y_{2}, z_{2},\cdots, x_{N}, y_{N}, z_{N})
       \end{equation}
and the sample {\bf x}$^{(1)}$, $\dots$, {\bf x}$^{(L)}$  
is given by a MD trajectory of finite length $L$,
see e.g. Refs.~\cite{garcia,amadei93}.
Before performing the PCA itself the data set must be corrected
for overall translation and rotation.
After centering according to Eq.~\ref{center} the overall
rotation of the considered system is subtracted. This can be
done by rotating each configuration to a reference
configuration by a least squares fit~\cite{MCLACH79}. 
This translation-rotation correction
will thus provide six eigenvalues equal to zero, which do not refer to
the purely internal motion.

Since the eigenvectors of the covariance matrix are mutually
orthogonal, i.e. uncorrelated, the motions in
the direction of each component, i.e. along the corresponding
eigenvector, can be interpreted as a ``mode''.
To visualize the motion of each mode one has to calculate the
coordinates including the mean of $\mathbf{x}$ and the
fluctuations of the positions described by the principal
component under consideration
       \begin{equation}
         \mathbf{x}_{i}^{(l)}= \tilde m_{i}^{(l)}\mathbf{m}_i +<\mathbf{x}>
       \end{equation}
where  the index $i$ denotes the $i$-th principal component and
       $l$ the $l$-th timestep of the trajectory of total length $L$. Note that this
       yields a ``caricature'' of an important dynamical motif hidden
       in the full dynamics~\cite{garcia}.

For the PCA carried out here, all
heavy atoms of the peptide backbone were included.
In addition, a united--atom representation
was introduced for each side chain by assigning a mass equivalent 
to that of all its constituting atoms which was located at its 
center of mass.
This procedure effectively reduces the number of atoms
to   $N=38$ and thus simplifies the analysis and its 
representation.
In line with previous findings~\cite{comp_ess}
it was confirmed that
this approximation does not lead to any significant alteration of the results.
Tests indicate that PCA including all atoms provides an
essentially identical picture for the low--order modes that are of interest.
The dependence of the resulting motions upon 
trajectory length was carefully evaluated and results
were found to be well converged for simulations 
on the order of 20~ns in duration.
For example, the dot product between the
slowest motion, i.e. $\{ {\bf m}_1 \}$ computed over
two distinct 20~ns trajectories and that from 
a 30-40~ns trajectory at the lowest temperature, 280~K,  
is about 0.98 therefore indicating
that the modes are essentially converged.
Finally, the modes discussed in this investigation are
found to exist also when using the zwitterionic octapeptide
and in the simulations with a larger solvation sphere around the peptide
as discussed in Appendix~\ref{app:sim}.

\subsection{Hydrogen Bond Analysis}
\label{app:HBA}

Concerning the definition of HBs in water, and thus the population
variable $h(t)$, 
we employ a standard structural criterion based on both the distance
($R_{\rm OH} < 2.6$~{\AA} as obtained from including the
entire first nearest neighbor peak in the intermolecular OH
radial distribution function)
and the angle, 
$\angle_{\rm O-H\cdots O} >130^\circ$ for $h(t)=1$~\cite{FHS,SNS,chandler96a,chandler96b,luzar00,BJB}.
As a check we also performed the analysis with only
the distance criterion 
and yet again with the distance criterion
but using a tighter angular criterion of
$\angle_{\rm O-H\cdots O} >150^\circ$.
Consistent 
with previous findings~\cite{SNS,BJB} the {\em absolute}
value of rate constants, $k_w$ and $\tilde k_w$,
is sensitive 
to the actual choice of the parameters of the definition and
was found to vary as much as a factor of 2-3 between different definitions.
However, we always obtained the same qualitative trends
that both HB numbers and rate constants 
($k_w$ and $\tilde k_w$) feature
a linear behavior with respect to temperature.
Similarly, 
for peptide/peptide and peptide/water HBs
the same criterion based on distances
($R_{\rm AH} < 2.6$~{\AA} where the proton acceptor A is either an O atom 
of a water molecule or of a peptide carbonyl group)
and angles
($\angle_{\rm D-H\cdots A} >130^\circ$
with the proton donor D being either water O or peptide 
backbone N) was used.
As a check we varied the distance to $R_{\rm AH} < 2.4$~{\AA} 
keeping the angle cutoff unaltered
and performed the analysis with only
the distance criterion.
Again we found the same general behavior of two temperature
regimes based on the dynamics of peptide/water HBs 
and the temperature dependence of $n_{\rm pp}$ and $n_{\rm bw}$
to be preserved. 
In conclusion,
although the HB definitions do change
the reported {\em absolute values} of the HB numbers and rate constants
as expected~\cite{SNS,BJB},
the qualitative behaviour remains the same and thus 
the trends in the temperature dependence for these 
quantities are not affected.

\newpage

%%%%%%%%%%%%%%%%%%%%%%%%%%%%%%%%%%%%%%%%%%%%%%%%%%%%%%%%%%%%%%%%%%%%%%%%%%%

%%%%%%%                       REFERENCES
%%%%%%%%%%%%%%%%%%%%%%%%%%%%%%%%%%%%%%%%%%%%%%%%%%%%%%%%%%%%%%%%%%%%%%%%%%%

%neccessary for the style file
\nocite{TitlesOn}
\bibliography{elastin_refs}

\begin{thebibliography}{}

\bibitem{ciba}
Pasquali-Ronchetti, I., Fornieri, C., Baccarani-Contri, M., \& Quaglino, D.
  (1995).
\newblock Ultrastructure of elastin.
\newblock In: {\em The molecular biology and pathology of elastic tissues} pp.
  31--42, Ciba Foundation ~: John Wiley \& Sons Ltd.

\bibitem{urry-jpcB}
Urry, D. (1997).
\newblock Physical chemistry of biological free energy transduction as
  demonstrated by elastic protein-based polymers.
\newblock {\em Journal of Physical Chemistry B, } {\bf 101}, 11007--11028.

\bibitem{tamburro99}
Debelle, L. \& Tamburro, A. (1999).
\newblock Elastin: molecular description and function.
\newblock {\em The International Journal of Biochemistry {\&} Cell Biology, }
  {\bf 31}, 261--272.

\bibitem{alix99}
Debelle, L. \& Alix, A. (1999).
\newblock The structures of elastins and their function.
\newblock {\em Biochimie, } {\bf 81}, 981--994.

\bibitem{Rees01}
Reiersen, H. \& Rees, A. (2001).
\newblock The hunchback and its neighbours: proline as an environmental
  modulator.
\newblock {\em Trends in Biochemical Sciences, } {\bf 26}, 679--684.

\bibitem{Tamburro02}
Martino, M., Perri, T., \& Tamburro, A. (2002).
\newblock Biopolymers and biomaterials based on elastomeric proteins.
\newblock {\em Molecular Bioscience, } {\bf 2}, 319--328.

\bibitem{urry02}
Urry, D., Hugel, T., Seitz, M., Gaub, H., Sheiba, L., Dea, J., Xu, J., \&
  Parker, T. (2002).
\newblock Elastin: a representative ideal protein elastomer.
\newblock {\em Phil. Trans. R. Soc. Lond. B, } {\bf 357}, 169--184.
\newblock doi:10.1098/rstb.2001.1023.

\bibitem{urry-angew}
Urry, D. (1993).
\newblock Molecular machines: how motion and other functions of living
  organisms can result from reversible chemical changes.
\newblock {\em Angewandte Chemie International Edition In English, } {\bf 32},
  819--841.

\bibitem{rees98}
Reiersen, H., Clarke, A., \& Rees, A. (1998).
\newblock Short elastin--like peptides exhibit the same temperature--induced
  structural transitions as elastin polymers: implications for protein
  engineering.
\newblock {\em Journal of Molecular Biology, } {\bf 283}, 255--264.

\bibitem{cryst1}
Urry, D., Long, M., \& Sugano, H. (1978).
\newblock Cyclic analog of elastin polyhexapeptide exhibits an inverse
  temperature transition leading to crystallization.
\newblock {\em Journal of Biological Chemistry, } {\bf 253}, 6301--6302.

\bibitem{cryst2}
Cook, W., Einspahr, H., Trapane, T., Urry, D., \& Bugg, C. (1980).
\newblock Crystal structure and conformation of the cyclic trimer of a repeat
  pentapeptide of elastin, {cyclo-(L-valyl-L-prolylglycyl-L-valylglycyl)$_3$}.
\newblock {\em Journal of American Chemical Society, } {\bf 102}, 5502--5505.

\bibitem{Chilkoti}
Nath, N. \& Chilkoti, A. (2001).
\newblock Interfacial phase transition of an environmentally responsive elastin
  biopolymer adsorbed on functionalized gold nanoparticles studied by colloidal
  surface plasmon resonance.
\newblock {\em Journal of American Chemical Society, } {\bf 123}, 8197--8202.
\newblock doi:10.1021/ja015585r.

\bibitem{Partidge62}
S.{\,}Partridge (1962).
\newblock Elastin.
\newblock {\em Advances in Protein Chemistry, } {\bf 17}, 227--302.

\bibitem{daggett01}
Li, B., Alonso, D., \& Daggett, V. (2001).
\newblock The molecular basis for the inverse temperature transition of
  elastin.
\newblock {\em Journal of Molecular Biology, } {\bf 305}, 581--592.
\newblock doi:10.1006/jmbi.2000.4306.

\bibitem{flory}
Hoeve, C. \& Flory, P. (1974).
\newblock The elastic properties of elastin.
\newblock {\em Biopolymers, } {\bf 13}, 677--686.

\bibitem{wasserman90}
Wasserman, Z. \& Salemme, F. (1990).
\newblock A molecular dynamics investigation of the elastomeric restoring force
  in elastin.
\newblock {\em Biopolymers, } {\bf 29}, 1613--1631.

\bibitem{daggett02c}
Li, B. \& Daggett, V. (2002).
\newblock Molecular basis for the extensibility of elastin.
\newblock {\em Journal of Muscle Research and Cell Motility, } {\bf 23},
  561--573.

\bibitem{NMR1}
Fleming, W., Sullivan, C., \& Torchia, D. (1980).
\newblock Characterization of molecular motions in {$^{13}$C}-labeled aortic
  elastin by {$^{13}$C-$^1$H} magnetic double resonance.
\newblock {\em Biopolymers, } {\bf 19}, 597--617.

\bibitem{NMR2}
Torchia, D. \& Piez, K. (1973).
\newblock Mobility of elastin chains as determined by {$^{13}$C} nuclear
  magnetic resonance.
\newblock {\em Journal of Molecular Biology, } {\bf 76}, 419--424.

\bibitem{biref}
Aaron, B. \& Gosline, J. (1980).
\newblock Optical properties of single elastin fibres indicate random protein
  conformation.
\newblock {\em Nature, } {\bf 287}, 865--867.

\bibitem{urry83}
Urry, D., Trapane, T., Long, M., \& Prasad, K. (1983).
\newblock Test of the librational entropy mechanism of elasticity of
  polypentapeptide of elastin.
\newblock {\em J.Chem.Soc., Faraday Trans.~1, } {\bf 79}, 853--868.

\bibitem{NMR3}
O.~Arad, M.~G. (1990).
\newblock Depsipeptide analogues of elastin repeating sequences: conformational
  analysis.
\newblock {\em Biopolymers, } {\bf 29}, 1651--1668.

\bibitem{urry85}
Urry, D., Shaw, R., \& Prasad, K. (1985).
\newblock Polypentapeptide of elastin: temperature dependence of ellipticity
  and correlation with elastomeric force.
\newblock {\em Biochemical and Biophysical Research Communications, } {\bf
  130}, 50--57.

\bibitem{joint}
Schreiner, E., Nicolini, C., Ludolph, B., Ravindra, R., Otte, N., Kohlmeyer,
  A., Rousseau, R., Winter, R., \& Marx, D. (2003).
\newblock Folding and unfolding of an elastin--like oligopeptide: ``inverse
  temperature transition'', re--entrance, and hydrogen--bond dynamics.
\newblock {\em submitted, } {\bf }.

\bibitem{winter}
Nicolini, C., Ravindra, R., Ludolph, B., \& Winter, R. (2003).
\newblock Characterization of the temperature- and pressure-induced inverse and
  re-entrant transition of the minimum elastin-like polypeptide {GVG(VPGVG)} by
  {DSC}, {PPC}, {CD} and {FT-IR} spectroscopy.
\newblock {\em Journal of Molecular Biology, } {\bf 0}, 0--0.
\newblock previous paper.

\bibitem{debelle95}
Debelle, L., Alix, A., Jacob, M., Huvenne, J., Berjot, M., Sombret, B., \&
  Legrand, P. (1995).
\newblock Bovine elastin and kappa-elastin secondary structure determination by
  optical spectroscopies.
\newblock {\em Journal of Biological Chemistry, } {\bf 270}, 26099--26103.

\bibitem{perry02}
Perry, A., Stypa, M., Foster, J., \& Kumashiro, K. (2002).
\newblock Observation of the glycines in elastin using {$^{13}$C} and
  {$^{15}$N} solid-state {NMR} spectroscopy and isotopic labeling.
\newblock {\em Journal of American Chemical Society, } {\bf 124}, 6832--6833.
\newblock doi:10.1021/ja017711x.

\bibitem{Kurkova}
Kurkov\'{a}, D., K\v{r}\'{\i}\v{z}, J., Schmidt, P., Dybal, J.,
  Rodr\'{\i}guez-Cabello, J., \& Alonso, M. (2003).
\newblock Structure and dynamics of two elastin-like polypentapeptides studied
  by {NMR} spectroscopy.
\newblock {\em Biomacromolecules, } {\bf 4}, 589--601.
\newblock doi:10.1021/bm025618a.

\bibitem{daggett02}
Li, B., Alonso, D., Benion, B., \& Daggett, V. (2001).
\newblock Hydrophobic hydration is an important source of elasticity in
  elastin-based biopolymers.
\newblock {\em Journal of American Chemical Society, } {\bf 123}, 11991--1198.
\newblock doi:10.1021/ja010363e.

\bibitem{daggett02b}
Li, B., Alonso, D., \& Daggett, V. (2002).
\newblock Stabilization of globular proteins via introduction of
  temperature-activated elastin-based switches.
\newblock {\em Structure, } {\bf 10}, 989--998.

\bibitem{urry85a}
Urry, D., Trapane, T., Iqbal, M., Venkatachalam, C., \& Prasad, K. (1985).
\newblock Carbon-13 {NMR} relaxation studies demonstrate an inverse temperature
  transition in the elastin polypentapeptide.
\newblock {\em Biochimie, } {\bf 24}, 5182--5189.

\bibitem{savitzky}
Savitzky, A. \& Golay, M. (1964).
\newblock Smoothing and differentiation of data by simplified least squares
  procedures.
\newblock {\em Analytical Chemistry, } {\bf 36}, 1627--1639.

\bibitem{soper}
Dixit, S., Crain, J., Poon, W., Finney, J., \& Soper, A. (2002).
\newblock Molecular segregation observed in a concentrated alcohol-water
  solution.
\newblock {\em Nature, } {\bf 416}, 829.
\newblock doi:10.1038/416829a.

\bibitem{garcia}
Garc\'{i}a, A.~E. (1992).
\newblock Large-amplitude nonlinear motions in proteins.
\newblock {\em Physical Review Letters, } {\bf 68}, 2696--2699.
\newblock doi:10.1103/PhysRevLett.68.2696.

\bibitem{amadei93}
Amadei, A., Linssen, A., \& Berendsen, H. (1993).
\newblock Essential dynamics of proteins.
\newblock {\em Proteins: Structure, Function and Genetics, } {\bf 17},
  412--425.

\bibitem{ica}
Hyv\"arinen, A., Karhunen, J., \& Oja, E. (2001).
\newblock {\em Independent component analysis} chapter~6.
\newblock John Wiley \& Sons.

\bibitem{schlitter93}
Schlitter, J. (1993).
\newblock Estimation of absolute and relative entropies of macromolecules using
  the covariance matrix.
\newblock {\em Chemical Physics Letters, }, {\bf 215}, 617--621.
\newblock doi:10.1016/0009-2614(93)89366-P.

\bibitem{karplus01}
Andriciolaei, I. \& Karplus, M. (2001).
\newblock On the calculation of entropy from covariance matrices of the atomic
  fluctuations.
\newblock {\em Journal of Chemical Physics, }, {\bf 115} (14), 6289--6292.
\newblock doi:10.1063/1.1401821.

\bibitem{FHS}
Stillinger, F.~H. (1975).
\newblock Theory and molecular models for water.
\newblock {\em Advances in Chemical Physics, }, {\bf 31}, 1--101.

\bibitem{chandler96a}
Luzar, A. \& Chandler, D. (1996).
\newblock Effect of environment on hydrogen bond dynamics in liquid water.
\newblock {\em Physical Review Letters, }, {\bf 76}, 928--931.
\newblock doi:10.1103/PhysRevLett.76.928.

\bibitem{chandler96b}
Luzar, A. \& Chandler, D. (1996).
\newblock Hydrogen bond kinetics in liquid water.
\newblock {\em Nature, }, {\bf 379}, 55--57.

\bibitem{SNS}
Starr, F., Nielsen, J., \& Stanley, H. (1999).
\newblock Fast and slow dynamics of hydrogen bonds in liquid water.
\newblock {\em Physical Review Letters, }, {\bf 82}, 2294--2297.
\newblock doi:10.1103/PhysRevLett.82.2294.

\bibitem{AC}
Chandra, A. (2000).
\newblock Effects of ion atmosphere on hydrogen-bond dynamics in aqueous
  electrolyte solutions.
\newblock {\em Physical Review Letters, }, {\bf 85}, 768--771.
\newblock doi:10.1103/PhysRevLett.85.768.

\bibitem{BJB}
Xu, H. \& Berne, B. (2001).
\newblock Hydrogen-bond kinetics in the solvation shell of a polypeptide.
\newblock {\em Journal of Physical Chemistry B, }, {\bf 105}, 11929--11932.
\newblock doi:10.1021/jp012749h.

\bibitem{tobias99}
Tarek, M. \& Tobias, D. (1999).
\newblock Environment dependence of the dynamics of protein hydration water.
\newblock {\em Journal of American Chemical Society, }, {\bf 121}, 9740--9741.
\newblock doi:10.1021/ja990643i.

\bibitem{tobias02a}
Tarek, M. \& Tobias, D. (2002).
\newblock Role of protein-water hydrogen bond dynamics in the protein dynamical
  transition.
\newblock {\em Physical Review Letters, }, {\bf 88}, 138101--1--138101--4.
\newblock doi:10.1103/PhysRevLett.88.138101.

\bibitem{tobias02b}
Tarek, M. \& Tobias, D. (2002).
\newblock Single-particle and collective dynamics of protein hydration water: a
  molecular dynamics study.
\newblock {\em Physical Review Letters, }, {\bf 89}, 275501--1--275501--4.
\newblock doi:10.1103/PhysRevLett.89.275501.

\bibitem{luzar00}
Luzar, A. (2000).
\newblock Extent of inter-hydrogen bond correlations in water. temperature
  effect.
\newblock {\em Chemical Physics, }, {\bf 258}, 267--276.

\bibitem{VanDerSpoel}
van~der Spoel, D., van Maaren, P., \& Berendsen, J. (1998).
\newblock A systematic study of water models for molecular simulation:
  derivation of water models optimized for use with a reaction field.
\newblock {\em Journal of Chemical Physics, }, {\bf 108}, 10220--10230.

\bibitem{stochastic}
{Brooks III}, C. \& Karplus, M. (1983).
\newblock Deformable stochastic boundaries in molecular dynamics.
\newblock {\em Journal of Chemical Physics, }, {\bf 79}, 6312--6325.

\bibitem{charmm22}
{MacKerell Jr.}, A.~D. {\em et~al.} (1998).
\newblock All-atom empirical potential for molecular modeling and dynamics
  studies of proteins.
\newblock {\em Journal of Physical Chemistry B, }, {\bf 102}, 3586--3616.

\bibitem{tip3p}
Jorgensen, W., Chandrasekhar, J., Madura, J., Impey, R., \& Klein, M. (1983).
\newblock Comparison of simple potential functions for simulating liquid water.
\newblock {\em Journal of Chemical Physics, }, {\bf 79}, 926--935.

\bibitem{ego1}
Eichinger, M., Heller, H., \& Grubm\"{u}ller, H. (2000).
\newblock {EGO}~-- an efficient molecular dynamics program and its application
  to protein dynamics simulations.
\newblock In: {\em Molecular Dynamics on Parallel Computers}, (Esser, R.,
  Grassberger, P., Grotendorst, J., \& Lewerenz, M., eds), pp. 154--174,
  Singapore: World Scientific.

\bibitem{ego2}
Eichinger, M., Grubm\"{u}ller, H., Heller, H., \& Tavan, P. (1997).
\newblock {FAMUSAMM}: An algorithm for rapid evaluation of electrostatic
  interactions in molecular dynamics simulations.
\newblock {\em Journal of Computational Chemistry, } {\bf 18}, 1729--1749.

\bibitem{PTdiagram}
Tamura, T., Yamaoka, T., Panitch, A., \& Tirrell, D. (2000).
\newblock Effects of temperature and pressure on the aggregation properties of
  an engineered elastin model polypeptide in aqueous solution.
\newblock {\em Biomacromolecules, } {\bf 1}, 552--555.
\newblock doi:10.1021/bm005606u.

\bibitem{MCLACH79}
McLachlan, A.~D. (1979).
\newblock Gene duplications in the structural evolution of chymotrypsin.
\newblock {\em Journal of Molecular Biology, } {\bf 128}, 49--77.

\bibitem{comp_ess}
Aalten, D.~V., Groot, B.~D., Findlay, J., Berendsen, H., \& Amadei, A. (1997).
\newblock A comparison of techniques for calculating protein essential
  dynamics.
\newblock {\em Journal of Computational Chemistry, } {\bf 18}, 169--181.
\newblock
  doi:10.1002/(SICI)1096-987X(19970130)18:2$<$169::AID-JCC3$>$3.0.CO;2-T.

\end{thebibliography}
\bibliographystyle{jmb-unsrt}

%%%%%%%%%%%%%%%%%%%%%%%%%%%%%%%%%%%%%%%%%%%%%%%%%%%%%%%%%%%%%%%%%%%%%%%%%%%
%%%%%%%                       Figures
%%%%%%%%%%%%%%%%%%%%%%%%%%%%%%%%%%%%%%%%%%%%%%%%%%%%%%%%%%%%%%%%%%%%%%%%%%%
\newpage

%%%%%%%%%%%%%%%%%%%%%%%%%%%%%%%%%%%%%%%%%%%%%%%%%%%%%%%%%%%%%%%%%%%%%%%%%%%
%%%%%%%                       Tables
%%%%%%%%%%%%%%%%%%%%%%%%%%%%%%%%%%%%%%%%%%%%%%%%%%%%%%%%%%%%%%%%%%%%%%%%%%%
\newpage
\end{document}